\documentclass[prd, aps, 
eqsecnum,
preprint,
amsmath,
amssymb]{revtex4-2}

\usepackage{multirow}
\usepackage[utf8]{inputenc}
\usepackage[T1]{fontenc}
\usepackage{physics}
\usepackage{natbib}
\usepackage{amsmath}
\usepackage{braket}
\usepackage{pgfplots}
\usepackage{subfig}
\usepackage{commath}
\usepackage{feynmp-auto}
\usepackage{bm}
\usepackage{slashed}
\usepackage{bbold}
\usepackage{tensor}
\usepackage{amssymb}
\usepackage{mathrsfs}
\usepackage{mathtools}
\usepackage{graphicx}
\usepackage{xspace}
\usepackage[dvipsnames]{xcolor}
\usepackage{ulem}
\usepackage{hyperref}

\graphicspath{ {./p/} }

\allowdisplaybreaks

\pgfplotsset{compat=1.18} 

\begin{document}


\title{Quark mixing and CP violation due to torsional interaction}

\author{Arnab Chakraborty}
\email{arnabchakraborty233@bose.res.in}
\author{Amitabha Lahiri}
\email{amitabha@bose.res.in}
\affiliation{Satyendra Nath Bose National Centre for Basic Sciences\\
	JD Block, Sector 3, Salt Lake, Kolkata 700106}

\begin{abstract}
The dynamics of fermions in spacetime generates torsion, ultimately leading to a current-current interaction independent of gravity. This torsional four-fermion interaction is generically chiral and non-universal and may have observable consequences. Furthermore, the fermion basis in which this torsional interaction is diagonal may be different from the mass basis and the flavor basis. We work out the structure of the quark sector, where  new CKM-like matrices  appear as a result of diagonalizing the torsional interaction. CP violation is modified in general due to the appearance of additional phases. We also consider various processes which are affected by it such as kaon decays in particular and show that even when the torsional interaction is taken to be diagonal in the flavor basis, flavor-changing processes are still affected due to the non-universality of the torsional coupling constants. Finally, we discuss rephasing invariants in the presence of the torsional interaction and show how the Jarlskog invariant is modified even when the torsion basis is aligned with the flavor basis.
\end{abstract}

\maketitle

\newpage

{\section{Introduction}}\label{intro}
The Standard Model (SM) of particle physics has been extremely successful in explaining a wide variety of experimental data. Nonetheless, it suffers from many well-known shortcomings which indicate a need for Beyond the Standard Model (BSM) physics. One important issue in the quark sector where observations appear to be in tension with the SM concerns the matter-antimatter asymmetry~\cite{Riotto:1999yt,Trodden:1998ym,Dine:2003ax,Morrissey:2012db,Canetti:2012zc}. The Cabibo-Kobayashi-Maskawa (CKM) matrix is the only known source of CP violation in the SM, but the predicted CP violation is insufficient to explain the observed baryon asymmetry of the Universe. Popular BSM approaches to resolve this tension require new symmetries like Supersymmetry~\cite{Carena:1997gx,Dine:2003ax} and/or new particles such as heavy sterile neutrinos, additional Higgs particles, or new gauge bosons~\cite{Akhmedov:1998qx,Cline:2011mm,Enomoto:2021dkl,Fukuda:2024pkh}. However, there is a type of low energy physics which does not receive the attention it deserves in particle physics, and that is interactions arising from the geometrical structure of spacetime. Although gravitation is considered to be unimportant in particle interactions, being extremely weak, except possibly in situations where quantum gravity effects could become relevant, spacetime torsion may not be as weak and could play an important role in particle physics. 

In this paper we consider a BSM interaction arising from spacetime torsion generated by the dynamics of fermions in curved spacetime. This interaction does not require new particles, nor is it based on any new symmetry, but arises naturally when fermions are put on a curved spacetime and contributes to CP violation and other physical phenomena. As we will shortly see, the torsion produces a four-fermion interaction  which involves all fermion species and is generically chiral and non-universal. The coupling constants are arbitrary, in the sense that they cannot be fixed \textit{a priori} from theoretical considerations alone but must be treated as undetermined parameters till when they are fixed by precision experiments. This interaction has been studied recently in the context of neutrino oscillations, where it was found that known observational results allowed for a torsional interaction one or two orders of magnitudes smaller than weak interactions. We will briefly review the origin of the torsional four-fermion interaction in Sec.~\ref{origin}. 

In this paper our focus is on the quark sector and how the torsional interaction affects the mixing between quarks of different flavors. It should be obvious that mixing between different generations of quarks in the presence of such an interaction results in flavor-changing neutral current (FCNC)-like processes. In general, the basis of fermion fields in which the torsional interaction is diagonal need not be the same as the flavor basis or the mass basis. This distinctness leads to a new set of CKM-like mixing matrices involving several new angles and phases, as we display in Sec.~\ref{quarks}. We assume that the left chiral doublets remain doublets, i.e., torsion does not explicitly break the continuous symmetries of the SM.

The new mixing matrices modify amplitudes for various processes. While tree-level FCNC processes are forbidden in the SM, the torsional interaction results in FCNC-like diagrams at the tree level for non-leptonic decay processes. We calculate the operators which appear from the torsional interaction in the case of the $K \to \pi\pi$ decay in Sec.~\ref{Processes}. Operators for some other flavor-changing processes are given in Appendix~\ref{operators}.

An important motivation for investigating torsional quark mixing was to look for new sources of CP violation. When the torsion basis is not aligned with the flavor basis, it is obvious that the mixing matrices between these bases will provide new sources of CP violation. We calculate the new rephasing invariants in Sec.~\ref{Jarlskog}. We also find that even when the  torsional interaction is diagonal in the flavor basis, it still contributes to CP violation.

A measure of CP violation in the SM is the Jarlskog invariant. In the quark sector it is constructed out of the elements of the CKM matrix. In the presence of torsion, there are new patterns of quark mixing, so the Jarlskog invariant could be affected. We take a Fierz rearrangement of the torsional interaction in Sec.~\ref{Revisited} to see how both the up-type and down-type quarks mix due to the presence of the new mixing matrices. We find that the Jarlskog invariant is modified even when the  torsional interaction is diagonal in the flavor basis.  

Torsional mixing of quarks leads to many interesting phenomena, of which we have discussed only a few examples in this paper. We end the paper by discussing other implications of it, including other examples of hadronic processes, as well as possible avenues of future investigations, in Sec.~\ref{summary}.

\section{Origin of the geometrical four-fermion interaction}\label{origin}
Let us briefly go over the origin of the four-fermion interaction from the dynamics of fermions in curved spacetime. We use the Einstein-Cartan-Sciama-Kibble (ECSK) formalism -- a framework particularly well-suited for describing fermions on a dynamical spacetime~\cite{Cartan:1923zea, Cartan:1924yea, Gasperini:2013cru, Gursey:1957, Finkelstein:1960, Hammond:2002rm, Hehl:1971qi, Hehl:1974cn, Hehl:1976kj, Hehl:2007bn, Kibble:1961ba, Mielke:2017nwt, Sciama:1964wt, Poplawski:2009fb, Chakrabarty:2018ybk}. This is essentially a way of writing gravity as an SO(3,1) gauge theory, for which the `internal space' is flat Minkowski space, identified with the tangent space at each point. A set of four 
vector fields $e^\mu_a$ called tetrads are defined on spacetime so that they span the internal space at each point. Their inverses $e_\mu^a$  are called co-tetrads and 
satisfy 
\begin{equation}
    g_{\mu\nu}e^\mu_a e^\nu_b = \eta_{ab}\,, \qquad \eta_{{ab}}e^a_\mu e^b_\nu = g_{\mu\nu}\,,
\end{equation}
where $g_{\mu\nu}$ is the metric on the curved spacetime and $\eta_{ab}$ is the flat Minkowski metric. The determinant of the co-tetrad viewed as a $4 \times 4$ matrix is related to the metric determinant by $|e| = \sqrt{|g|}$. 
The spin connection $A_\mu{}^{ab}$ is needed to define covariant derivatives of objects with indices in the internal space. 
The tetrads are connection-compatible, leading to a relation between the familiar connection components and the spin connection,
\begin{equation}\label{Gamma-spin}
    \nabla_\mu e^\lambda_a 
    = 0\, \quad \Rightarrow \quad \Gamma^{\lambda}{}_{\mu\nu} = e^\lambda_a\partial_\mu e_\nu^a + A_{\mu}{}^{b}{}_{a} e^\lambda_be_\nu^a\,. 
\end{equation}
Then the field strength of the spin connection  $F_{\mu\nu}{}^{ab}(A) = \partial_\mu A_\nu{}^{ab} - \partial_\nu A_\mu{}^{ab} + A_{\mu}{}^{a}{}_{c} A_{\nu}{}^{cb} -  A_{\nu}{}^{a}{}_{c} A_{\mu}{}^{cb}\,$, is related to the Ricci scalar constructed with $\Gamma^{\lambda}{}_{\mu\nu}$ as $R = F_{\mu\nu}{}^{ab}(A) e^\mu_a e^\nu_b.$

For the covariant derivative of spinor fields, we start by defining constant $\gamma^a$ matrices on the internal flat space,
satisfying $\left[\gamma^a, \gamma^b\right]_{+} = 2\eta^{ab}$\,. Then the covariant derivative
of fermion fields is $D_\mu\psi = \partial_\mu\psi - \frac{i}{4} A_\mu{}^{ab} \sigma_{ab}\psi\,,$ where $\sigma_{ab} = \frac{i}{2}\left[\gamma_a\,, \gamma_b\right]_{-}$\,. The action for gravity plus fermions is thus given by
\begin{equation}\label{action.1}
	S = \int |e| d^4x \left(\frac{1}{2\kappa} F_{\mu\nu}{}^{ab}(A) e^\mu_a e^\nu_b +  \frac{1}{2} \left( i\bar{\psi} \slashed{D}\psi + h. c.\right)\right) \,,
\end{equation}
where $\kappa = 8\pi G.$ Here we have written $\slashed{D} = \gamma^{\mu}D_{\mu}=e^\mu_a \gamma^aD_\mu\,.$ 
%
 

It is convenient to split the spin connection as $A_\mu{}^{ab} = \omega_\mu{}^{ab} +\Lambda_\mu{}^{ab}\,$, where we have defined
\begin{equation}
    \omega_\mu{}^{ab} = e^m_\mu\eta^{ac}\eta^{bd}(\Omega_{mcd}-\Omega_{cdm}+\Omega_{dmc}),
\end{equation}
using the Ricci rotation coefficients $\Omega_{cd}{}^{a}= e^\mu_ce^\nu_d\partial_{[\mu}e^a_{\nu]} \equiv e^\mu_ce^\nu_d(\partial_{\mu}e^a_{\nu}-\partial_{\nu}e^a_{\mu})$. 
When this $\omega$ is inserted in place of the spin connection in Eq.~\eqref{Gamma-spin}, the resulting $\Gamma_{\mu\nu}^\lambda$ 
are the Christoffel symbols, which correspond to the torsionless Levi-Civita connection, while $\Lambda$ is called the contorsion field. 
The tetrads $e^\mu_a$ and the contorsion components $\Lambda_\mu{}^{ab}$ are the independent dynamical variables. In terms of $\omega$ and $\Lambda$, the field strength $F_{\mu\nu}$ can be written as 
\begin{equation}
    F_{\mu\nu}{}^{ab}[A] = F_{\mu\nu}{}^{ab}[\omega] + \partial_{[\mu}\Lambda_{\nu]}{}^{ab} + \omega_{[\mu}{}^{[a}{}_{\left|c\right|}\Lambda_{\nu]}{}^{\left|c\right|b]} + \Lambda_{[\mu}{}^{a}{}_{\left|c\right|}\Lambda_{\nu]}{}^{cb}\,. 
    \label{F[A]}
\end{equation}
%
%
The first term in Eq.~\eqref{F[A]} is simply the torsion-free Ricci scalar, $\widetilde R = e^\mu_a e^\nu_bF_{\mu\nu}{}^{ab}[\omega]$. The terms linear in the contorsion $\Lambda$ combine to make a total derivative  and can therefore ignored. Then the gravitational action becomes
%
\begin{equation}
S_G
=
\frac1{2\kappa}
\int d^4x\,|e|
\left[
\widetilde R
+
\Lambda_{abc}\Lambda^{bac}
-
\Lambda_a\Lambda^a
\right],
\end{equation}
where we have written $\Lambda_a{}^{bc}=e^\mu_a\Lambda_\mu{}^{bc}$ and $\Lambda^a=\Lambda_{b}{}^{ab}$. Similarly decomposing the fermionic covariant derivative into its torsion-free and contorsion pieces yields the ordinary Dirac action together with a linear coupling between the contorsion and a fermion current. The action of the fermionic fields is thus
\begin{equation}
S_\psi
=
\int d^4x\,|e|\,
\Big[\frac{i}{2}
\left(
\bar\psi\gamma^\mu \widetilde D_\mu\psi
-
(\widetilde D_\mu\bar\psi)\gamma^\mu\psi
\right)
+\frac{1}{8}
\bar\psi
\left[
\gamma^\mu,\sigma_{ab}
\right]_{+}
\psi\Lambda_\mu{}^{ab}
\,\Big],
\end{equation}
where $\widetilde D_\mu$ is the covariant derivative constructed from $\omega_\mu{}^{ab}$ alone. In this action, the independent fields are the tetrads and the contorsion.

Since the action does not contain derivatives of the contorsion field,  it is not a propagating field. Varying the total action with respect to $\Lambda$ produces an algebraic field equation 
\begin{equation}\label{contorsion-eqn}
\Lambda^{bac}-\eta^{bc}\Lambda^a=-\frac{\kappa}{4}\epsilon^{acbd}\bar\psi\gamma_d\gamma_5\psi,
\end{equation}
which shows that the contorsion is fully expressible in terms of the fermions.
Taking the trace of Eq.~\eqref{contorsion-eqn} with respect to the $b,c$ indices, we see that $\Lambda^a = 0$, so the solution for the contorsion is 
\begin{equation}
\Lambda_{cab}=-\frac{\kappa}{4}\epsilon_{abcd}\bar\psi\gamma^d\gamma_5\psi.
\end{equation}
Substituting this solution back into the action eliminates the contorsion and generates a local four-fermion interaction. The total action becomes
\begin{equation}
S
=
\int d^4x\,|e|
\left[
\frac{\widetilde R}{2\kappa}
+
\frac{i}{2}
\left(
\bar\psi\gamma^\mu\widetilde D_\mu\psi
-
(\widetilde D_\mu\bar\psi)\gamma^\mu\psi
\right)
+
\frac{3\kappa}{16}
\left(
\bar\psi\gamma_a\gamma^5\psi
\right)
\left(
\bar\psi\gamma^a\gamma^5\psi
\right)
\right].
\end{equation}
%

However, this is not the most general form of coupling between contorsion and fermions. We wrote the above action for a single species of fermions for simplicity. In a realistic theory however, the action must include all fermionic degrees of freedom of the Standard Model. Since the contorsion $\Lambda_\mu{}^{ab}$ is independent of the tetrads, there is no symmetry principle that requires its coupling to be universal across different fermion species, unlike the gravitational part of the spin connection represented by $\omega_\mu{}^{ab}.$ Thus the contorsion field generically couples to different fermion species with different strengths.
Moreover, the interaction of fermions with $\Lambda_\mu{}^{ab}$ does not rely on the existence of a fermion mass term and can therefore be present even for chiral fermions, such as a purely left-handed neutrino.  Another way of seeing this is that being a gauge field for the gauge theory of the Poincar\'e group~\cite{Hehl:2023khc, Ali:2009ee}, contorsion can couple independently to their left and right-chiral components since $\psi_L$ and $\psi_R$ belong to different representations of the Lorentz group and therefore transform independently under it. On the other hand the Levi--Civita connection $\omega_\mu{}^{ab}$, being a function of the tetrad fields, couples universally to all fermions irrespective of their chirality.  It is also clearly not necessary that the coupling of contorsion to the fermions be diagonal in either the flavor basis or the mass basis. Let us therefore define the torsion basis fields as the ones for which the coupling to the contorsion field is diagonal. They will be denoted with a superscript $\tau$ below as $\psi^{\tau}_{iL}$ and $\psi^{\tau}_{iR}$.

The fermionic Lagrangian in curved spacetime can therefore be written in the general form~\cite{Chakrabarty:2019cau}
\begin{align}
	{\mathscr L}_\psi &= \sum\limits_{i}	\left[\frac{i}{2}\bar{\psi}^{\tau}_i\gamma^\mu\partial_\mu\psi^{\tau}_i - 
	\frac{i}{2}\partial_\mu\bar{\psi}^{\tau}_i\gamma^\mu\psi^{\tau}_i 	- \frac{1}{4} \epsilon_{abcd}\omega_{\mu}{}^{ab} e^{\mu c}  \,
	\bar{\psi}^{\tau}_i \gamma^d\gamma^5 \psi^{\tau}_i\,\right. \notag \\ 
	&\qquad\qquad \qquad \qquad \left.
	- \frac{1}{4}\epsilon_{abcd} \Lambda_{\mu}{}^{ab} e^{\mu c}
	\left(\tilde{\lambda}_{iL}\bar{\psi}^{\tau}_{iL} \gamma^d \psi^{\tau}_{iL} + \tilde{\lambda}_{iR}\bar{\psi}^{\tau}_{iR} \gamma^d\psi^{\tau}_{iR}\right)
	\right]\,,\label{L_psi_all}
\end{align}
where the sum runs over all species of fermions. The curvature term in the action~(\ref{action.1}) remains as before, and we find 
\begin{equation}\label{chiral.torsion}
	\Lambda_{\mu}{}^{ab} = -\frac{\kappa}{4}\epsilon^{abcd}e_{c\mu} \sum\limits_i \left(\tilde{\lambda}_{iL}\bar{\psi}^{\tau}_{iL}\gamma_d \psi^{\tau}_{iL} + \tilde{\lambda}_{iR}\bar{\psi}^{\tau}_{iR}\gamma_d \psi^{\tau}_{iR}\right)\,.
\end{equation}
As before, we can put this solution back into the action to get a four-fermion interaction term~\cite{Chakrabarty:2019cau},
\begin{equation}\label{4fermi}
{\mathscr L}_{4\psi} =	-\frac{3\kappa}{16}\left(\sum\limits_i \left(\tilde{\lambda}_{iL}\bar{\psi}^{\tau}_{iL} \gamma_a \psi^{\tau}_{iL} + \tilde{\lambda}_{iR}\bar{\psi}^{\tau}_{iR} \gamma_a \psi^{\tau}_{iR}\right)\right)^2\,.
\end{equation}
We will call this the geometrical or torsional four-fermion interaction. This interaction term does not depend explicitly on either the background metric or the Levi--Civita connection $\omega_\mu{}^{ab}$ and therefore contributes to usual laboratory experiments where the spacetime is very close to flat. Strictly speaking, spacetime is never exactly flat in the presence of matter; however, for the matter densities encountered in terrestrial experiments, the resulting curvature is negligibly small and the Minkowski approximation is adequate. The four-fermion interaction itself also contributes to the energy--momentum tensor and thus sources gravity through the Einstein equations. The effect on the metric due to that is extremely small and has a negligible impact on the spacetime geometry. Consequently, fermions in ordinary experiments obey the flat-spacetime Dirac equation, modified by an additional cubic self-interaction term from the four-fermion interaction.


\section{Quarks and family symmetry breaking}\label{quarks}

{In this paper, our focus is on  how the geometrical interaction affects discrete symmetries in the quark sector. We start with the usual Standard Model setup by taking the quarks in the weak or flavor basis~\cite{Cheng:1984vwu,Buras:1998raa,Buras:2020xsm}. Our notation is similar to the one in~\cite{Buras:1998raa}. Then the left chiral quarks are weak doublets,} 
%
\begin{equation}\label{flavor.1}
Q_{AL} =\left\{\begin{pmatrix} u_L \\ d^{\prime}_L \end{pmatrix}, \begin{pmatrix} c_L \\ s^{\prime}_L \end{pmatrix}, \begin{pmatrix} t_L \\ b^{\prime}_L \end{pmatrix}\right\}\,,
\end{equation}
where $A=1,2,3$ is the generation index, while the right-handed quarks are weak singlets and denoted by the first generation label,
\begin{equation}\label{flavor.2} 
u_{AR}=\{u_R, c_R, t_R\}, \hspace{0.5cm} d_{AR}=\{d_R, s_R, b_R\}\,.
\end{equation}
The electroweak gauge interactions of the quarks are given by 
\begin{align}\label{gauge}
\mathscr{L}^{q}_{\text{gauge}}&=-\frac{g}{\sqrt{2}}\bar{u}_{AL}\slashed{W}^+d^{\prime}_{AL}-\frac{g}{\sqrt{2}}\bar{d}^{\prime}_{AL}\slashed{W}^-u_{AL}-\bar{u}_{AL}\Bigl(\frac{g'}{6}\slashed{B}+\frac{g}{2}\slashed{W}^3\Bigl)u_{AL}\notag \\
& \hspace{2cm} -\bar{d}^{\prime}_{AL}\Bigl(\frac{g'}{6}\slashed{B}-\frac{g}{2}\slashed{W}^3\Bigl)d^{\prime}_{AL}-\frac{2}{3}g'\bar{u}_{AR}\slashed{B}u_{AR}+\frac{1}{3}g'\bar{d}_{AR}\slashed{B}d_{AR}\,,
\end{align}
%
where the couplings $g$ and $g'$ are related by $e=g\sin{\theta_W}=g'\cos{\theta_W}$, with $\theta_W$ being the weak mixing angle. The first two terms of Eq.~(\ref{gauge}) correspond to the charged current interactions of the quarks.   

Quark masses come from the Yukawa interactions, given by
\begin{equation}\mathscr{L}_{\text{Yuk}}=-Y^{AB}_d\bar{Q}_{AL}\phi d_{BR}
     -Y^{AB}_u\bar{Q}_{AL}\widetilde{\phi} u_{BR}+h.c.
\end{equation}
 where $Y_u$ and $Y_d$ are complex matrices with constant entries, $\phi$ is the Higgs doublet
 \begin{equation}
 \phi(x)=\begin{pmatrix} \phi^+(x)\\ \frac{v+H(x)+i\zeta(x)}{\sqrt{2}} \end{pmatrix}\,,
 \end{equation} 
and $\widetilde{\phi}=i\sigma^2\phi^*$. Therefore the Lagrangian can be rewritten as
\begin{align}\label{action.3}\mathscr{L}_{\text{Yuk}} = & -\frac{v}{\sqrt{2}}Y^{AB}_d\bar{d}^{\prime}_{AL}d_{BR}-\frac{v}{\sqrt{2}}Y^{AB}_u\bar{u}_{AL}u_{BR}-Y^{AB}_d\bar{u}_{AL}\phi^+ d_{BR}+Y^{AB}_u\bar{d}^{\prime}_{AL}\phi^- u_{BR}\notag  \\ 
& \qquad \qquad -\frac{1}{\sqrt{2}}Y^{AB}_d\bar{d}^{\prime}_{AL}(H+i\zeta) d_{BR}-\frac{1}{\sqrt{2}}Y^{AB}_u\bar{u}_{AL}(H-i\zeta) u_{BR}+h.c.
\end{align}
The weak or flavor basis and the mass basis are connected via the CKM matrix $V$, 
\begin{equation}\label{CKM} 
\begin{pmatrix} d^{\prime}_L \\ s^{\prime}_L \\ b^{\prime}_L \end{pmatrix}=\begin{pmatrix} V_{ud} \hspace{.5cm} V_{us} \hspace{.5cm} V_{ub} \\ V_{cd} \hspace{.5cm} V_{cs} \hspace{.5cm} V_{cb} \\ V_{td} \hspace{.5cm} V_{ts} \hspace{.5cm} V_{tb} \end{pmatrix}\begin{pmatrix} d_L \\ s_L \\ b_L \end{pmatrix}\,.
\end{equation}
From the Yukawa interactions, after rotating to the mass basis via Eq.~(\ref{CKM}), we get the diagonalized quark mass terms 
\begin{equation}\label{mass-term}
\mathscr{L}_{\text{mass}}=-\frac{v}{\sqrt{2}}\bar{d}_{L}M_dd_{R}-\frac{v}{\sqrt{2}}\bar{u}_{L}M_uu_{R}+h.c.
\end{equation}
Transforming to the mass basis via Eq.~(\ref{CKM}), the charged current interactions of quarks can be written as  
\begin{equation}
\mathscr{L}^q_{\text{CC}}=-\frac{g}{\sqrt{2}}\bar{u}_{AL}\slashed{W}^+V_{AB} d_{BL}+h.c.
\end{equation} 
For the charged current interaction $\mathscr{L}^q_{\text{CC}}$ to be CP invariant, we need $V=V^*$ which does not hold for 3 or more generations of quarks. Thus the CKM matrix leads to CP violation in the standard model. The neutral current interaction is CP invariant in the standard model.

Let us assume that the coupling of quarks to contorsion does not break the electroweak SU(2) symmetry, as that would appear as an explicit symmetry breaking term in the electroweak sector. Then the left-chiral components of the torsion basis fields can be expressed in terms of the flavor basis fields as 
\begin{equation}\label{alignment}
u^{\tau}_{AL}=(V_1)_{AB}u_{BL}\,, \qquad d^{\tau}_{AL}=(V_1)_{AB}d^{\prime}_{BL}\,,
\end{equation} 
where the same U(3) matrix $V_1$ appearing in both the equations ensures the preservation of the SU(2) symmetry. We also have the right-chiral components given by
\begin{equation}
u^{\tau}_{AR}=(V_2)_{AB}u_{BR}\,, \qquad d^{\tau}_{AR}=(V_3)_{AB}d_{BR}\,,
\end{equation}
where $V_2$ and $V_3$ are also U(3) matrices.
Then the action for quark fields in curved spacetime, written in terms of the torsion basis fields, becomes  
%
\begin{equation}\label{action.2}
	 S = \int |e| d^4x \left(\frac{1}{2\kappa} F_{\mu\nu}{}^{ab}(A) e^\mu_a e^\nu_b +  \frac{1}{2} \left( i \bar{Q}^{\tau}_{AL}\slashed{D}Q^{\tau}_{AL}+i \bar{u}^{\tau}_{AR}\slashed{D}u^{\tau}_{AR}+i \bar{d}^{\tau}_{AR}\slashed{D}d^{\tau}_{AR} + h. c.\right) \right) \,.
\end{equation}
Let us write the torsional interaction of the quarks by splitting the spin-connection $A$ as we did in Sec.~\ref{origin}. We note that in usual laboratory experiments, the spacetime is very close to being flat. Of course, spacetime is never exactly flat in the presence of matter, but for the kind of matter densities which appear in laboratory experiments the deviation from flatness is negligible. So we can set $\omega_\mu^{ab} \approx 0$ and 
$e^\mu_a\approx \delta^\mu_a$ for such experiments, and we can use ordinary flat space quantum field theory to analyze the interactions. 
The geometrical four-quark interaction term that will contribute to the laboratory experiments is 
\begin{align}\label{quark-contorsion}
	{\mathscr L}_q &= \frac{1}{8}\sum\limits_A \left(\tilde{\lambda}^Q_{AL}\bar{Q} ^{\tau}_{AL}\slashed{\Lambda}^{ab} \sigma_{ab}Q ^{\tau}_{AL} + \tilde{\lambda}^u_{AR}\bar{u}^{\tau} _{AR} \slashed{\Lambda}^{ab} \sigma_{ab}u^{\tau}_{AR}+ \tilde{\lambda}^d_{AR}\bar{d}^{\tau} _{AR} \slashed{\Lambda}^{ab} \sigma_{ab}d ^{\tau}_{AR}\right)+ h. c.
\end{align}
Using the identity $\gamma^a \gamma^b \gamma^c=\eta^{ab}\gamma^c+\eta^{bc}\gamma^a-\eta^{ac}\gamma^b-i\,\epsilon^{abcd}\gamma_d\gamma^5$, this interaction can be written as  
\begin{align}
	{\mathscr L}_q &=-\frac{1}{4} \epsilon_{abcd}\Lambda^{cab}\sum\limits_A   \,
	\left(\tilde{\lambda}^Q_{AL}\bar{u}^{\tau}_{AL} \gamma^du^{\tau} _{AL}+ \tilde{\lambda}^Q_{AL}\bar{d}^{\tau}_{AL} \gamma^dd^{\tau}_{AL} + \tilde{\lambda}^u_{AR}\bar{u}^{\tau} _{AR} \gamma^du ^{\tau}_{AR}+ \tilde{\lambda}^d_{AR}\bar{d}^{\tau}_{AR} \gamma^dd^{\tau} _{AR}\right)\,
\end{align}
Varying the total action \eqref{action.2} with respect to the connection, we get an algebraic equation of motion for the contorsion, 
%
\begin{equation}
	\Lambda^{ab}_c = -\frac{\kappa}{4}\tensor{\epsilon}{^{ab}}{_{cd}}\,\sum\limits_A\left(\tilde{\lambda}^Q_{AL}\bar{u}^{\tau}_{AL} \gamma^du^{\tau} _{AL}+ \tilde{\lambda}^Q_{AL}\bar{d}^{\tau}_{AL} \gamma^dd^{\tau}_{AL} + \tilde{\lambda}^u_{AR}\bar{u}^{\tau} _{AR} \gamma^du ^{\tau}_{AR}+ \tilde{\lambda}^d_{AR}\bar{d}^{\tau}_{AR} \gamma^dd^{\tau} _{AR}\right) \,.
\end{equation}
Putting this back into the action we get a four-fermion interaction term,
\begin{equation}\label{torsional int.}
{\mathscr L}^q_{4\psi} = -\frac{1}{2}\left[\sum\limits_A\left({\lambda}^Q_{AL}\bar{u}^{\tau}_{AL} \gamma^du^{\tau} _{AL}+ {\lambda}^Q_{AL}\bar{d}^{\tau}_{AL} \gamma^dd^{\tau}_{AL} + {\lambda}^u_{AR}\bar{u}^{\tau} _{AR} \gamma^du ^{\tau}_{AR}+ {\lambda}^d_{AR}\bar{d}^{\tau}_{AR} \gamma^dd^{\tau} _{AR}\right)\right]^2\,,
\end{equation}
where we have rescaled the couplings as ${\lambda}_{AL,R}=\sqrt{\frac{3\kappa}{8}}\tilde{\lambda}_{AL,R}$. Although the factor $\sqrt{\kappa}$ appears explicitly in the definition of $\lambda_{AL,R}$, this does not imply that the couplings are necessarily of order $10^{-19}\,\mathrm{GeV}^{-1}$. The reason is that the dimensionless parameters $\tilde{\lambda}_{AL,R}$ are not fixed by the theory and may, in principle, take arbitrarily large values. 
(Arguments based on perturbation theory cannot constrain them to be small because, like its twin quantum gravity, the torsional interaction is not amenable to perturbative quantization. We will come back to this point later.) Consequently, the couplings $\lambda_{AL,R}$ cannot be predicted from theoretical considerations alone -- it is not possible to surmise even their order of magnitude. Instead, they should be regarded as phenomenological parameters whose values must ultimately be constrained by comparing theoretical predictions with experimental data. Since this interaction interaction originates from spacetime geometry, we will sometimes refer to it as a geometrical interaction.  We can rewrite~\eqref{torsional int.} in the flavor basis as 
\begin{equation}
{\mathscr L}^q_{4\psi}=-\frac{1}{2}\left[\bar{u}_{L} \gamma^d{\Lambda}^Q_{L}u_{L}+ \bar{d}^{\prime}_{L} \gamma^d{\Lambda}^Q_{L}d^{\prime}_{L}+\bar{u}_{R} \gamma^d{\Lambda}^u_{R}u_{R}+ \bar{d}_{R} \gamma^d{\Lambda}^d_{R}d_{R}\right]^2\,,
\end{equation}
where the torsional coupling matrices have been defined as 
\begin{subequations}
\begin{equation}\label{coupling}
{\Lambda}^Q_{L}=V_1^{\dag}\text{diag}({\lambda}^Q_{1L},{\lambda}^Q_{2L},{\lambda}^Q_{3L})V_1,
\end{equation}
\begin{equation}
{\Lambda}^u_{R}=V_2^{\dag}\text{diag}({\lambda}^u_{1R},{\lambda}^u_{2R},{\lambda}^u_{3R})V_2,
\end{equation}
\begin{equation}
{\Lambda}^d_{R}=V_3^{\dag}\text{diag}({\lambda}^d_{1R},{\lambda}^d_{2R},{\lambda}^d_{3R})V_3.
\end{equation}
\end{subequations} 
We see that now the geometrical interaction is no longer generation-diagonal as there is mixing among the various flavors due to the hermitian matrices ${\Lambda}^Q_{L}$, ${\Lambda}^u_{R}$ and ${\Lambda}^d_{R}$. As a result, FCNCs are possible at tree level due to the geometrical interaction.

Since the particles observed in experiments are states with definite masses, we have to rotate to the mass basis if we want to calculate experimental observables. The geometrical interaction written in the mass basis, after the rotation using Eq.~(\ref{CKM}), thus becomes
\begin{equation}\label{geom-action}
{\mathscr L}^q_{4\psi}=-\frac{1}{2}\left[\bar{u}_{L} \gamma^d {\Lambda}^Q_{L} u_{L}+ \bar{d}_{L} \gamma^dV^{\dag} {\Lambda}^Q_{L}  Vd_{L}+\bar{u}_{R} \gamma^d {\Lambda}^u_{R} u_{R}+ \bar{d}_{R} \gamma^d {\Lambda}^d_{R} d_{R}\right]^2\,.
\end{equation}
Let us rewrite Eq.~(\ref{geom-action}) with explicit family indices on the fields and matrices. This will be useful in parametrizing the mixing matrices $V$, $V_1$, $V_2$, $V_3$.
\begin{align}\label{geom-action.2}{\mathscr L}^q_{4\psi}&=-\frac{1}{2}\left[\sum\limits_{A,I,J}\left({\lambda}^Q_{AL}\bar{u}_{JL} (V_1^{\dag})_{JA}\gamma^d(V_1)_{AI} u_{IL} + {\lambda}^Q_{AL}\bar{d}_{JL} (V^{\dag}V_1^{\dag})_{JA}\gamma^d(V_1V)_{AI} d_{IL}\right.\right.\notag \\ &\qquad\qquad\left.\left.\hspace{2cm}+ {\lambda}^u_{AR}\bar{u}_{JR} (V_2^{\dag})_{JA}\gamma^d(V_2)_{AI} u_{IR} + {\lambda}^d_{AR}\bar{d}_{JR} (V_3^{\dag})_{JA}\gamma^d(V_3)_{AI} d_{IR}\right)\right]^2\,.
\end{align} 
We see that family mixing is now controlled by the four unitary matrices $V$, $V_1$, $V_2$, $V_3$. Of these, the form of the CKM matrix $V$ is known from the SM. It contains four free parameters while each of $V_1$, $V_2$, $V_3$ consists of six parameters. (How to get this number is shown in Appendix~\ref{counting}.) {There are also the nine $\lambda$ couplings, making a total of 27 new free parameters not present in the SM.} The CKM matrix $V$ is given in the standard parametrization as~\cite{ParticleDataGroup:2024cfk}  
\begin{equation}V=
\begin{pmatrix}
c_{12} c_{13} & s_{12} c_{13} & s_{13} e^{-i\delta} \\
- s_{12} c_{23} - c_{12} s_{23} s_{13} e^{i\delta} &
c_{12} c_{23} - s_{12} s_{23} s_{13} e^{i\delta} &
s_{23} c_{13} \\
s_{12} s_{23} - c_{12} c_{23} s_{13} e^{i\delta} &
- c_{12} s_{23} - s_{12} c_{23} s_{13} e^{i\delta} &
c_{23} c_{13}
\end{pmatrix}\,,
\end{equation} 
 with $c_{ij} \equiv \cos\theta_{ij}$, $s_{ij} \equiv \sin\theta_{ij}$. As explained in Appendix~\ref{counting}, the
matrices $V_1$, $V_2$, and $V_3$ can be parametrized as 
\begin{equation}
V_1=\tilde{V}_1 {D}^{(1)}\,, \qquad V_2=\tilde{V}_2 {D}^{(2)}\,, \qquad V_3=\tilde{V}_3 {D}^{(3)}\,,
\end{equation}
where the CKM-like matrices $\tilde{V}_k$ are given by
\begin{equation}
\tilde{V}_k=
\begin{pmatrix}
\cos a_k \cos b_k & \sin a_k \cos b_k & \sin b_k e^{-i\delta_k} \\
- \sin a_k \cos c_k - \cos a_k \sin c_k \sin b_k e^{i\delta_k} &
\cos a_k \cos c_k - \sin a_k \sin c_k \sin b_k e^{i\delta_k} &
\sin c_k \cos b_k \\
\sin a_k \sin c_k - \cos a_k \cos c_k \sin b_k e^{i\delta_k} &
- \cos a_k \sin c_k - \sin a_k \cos c_k \sin b_k e^{i\delta_k} &
\cos c_k \cos b_k
\end{pmatrix}
\end{equation}
for $k=1,2,3$, and we have introduced the notation $a=\theta_{12},b=\theta_{13},c=\theta_{23}$ for each $k$. The parameters in $\tilde{V}_k$ are the 3 mixing angles $a_k, b_k, c_k$ and the CKM-like phase $\delta_k$. The diagonal phase matrices ${D}^{(k)}$ are parametrized by phases $\alpha^{(k)}_1, \alpha^{(k)}_2 $ as
\begin{equation}
    {D}^{(k)}=\text{diag}(1,e^{i\alpha^{(k)}_1},e^{i\alpha^{(k)}_2}) \,.
\end{equation}
The phases $\delta_k, \alpha^{(k)}_1, \alpha^{(k)}_2$ are additional sources of CP violation beyond the SM. We can see that there will be a proliferation of terms due to a large number of free parameters if we want to find the contribution of the geometrical interaction to any process. In order to gain some insight into this interaction by calculating its effect on observables such as decay amplitudes and CP asymmetries, we will make a simplifying assumption to reduce the number of free parameters. We will consider the special case that the torsion and flavor bases are aligned, i.e., the $V_i$ matrices are diagonal. Then these are diagonal U(3) matrices, 
$V_i=\text{diag}\big(e^{i\zeta^{(i)}_1},e^{i\zeta^{(i)}_2},e^{i\zeta^{(i)}_3}\big)$, and the geometrical interaction in the mass basis in Eq.~(\ref{geom-action.2}) becomes
\begin{align}\label{geom-action.3}
{\mathscr L}^q_{4\psi}=&-\frac{1}{2}\left[\sum\limits_{A,I,J}\left({\lambda}^Q_{AL}\bar{u}_{AL} \gamma^d u_{AL} + {\lambda}^Q_{AL}\bar{d}_{JL} (V^{\dag})_{JA}\gamma^d(V)_{AI} d_{IL}
\right.\right.\notag \\ &\qquad\qquad\qquad\left.\left.\qquad\qquad
+ {\lambda}^u_{AR}\bar{u}_{AR} \gamma^d u_{AR} + {\lambda}^d_{AR}\bar{d}_{AR} \gamma^d d_{AR}\right)\right]^2\,.
\end{align}
Thus, we are left with only the CKM matrix $V$ in this scenario with a total of 4 free parameters and no extra parameter or CP violating phase appears beyond what is already present in the Standard Model. However, due to the presence of the couplings ${\lambda}^Q_{AL}$, the interaction term given in Eq.~(\ref{geom-action.3}) still causes the breaking of CP symmetry. Thus, CP violation still arises beyond the Standard Model due to the geometrical interaction.

\section{Processes}\label{Processes}

It is well-known that there are no flavor-changing neutral currents (FCNCs) at tree level in the Standard Model as a result of the unitarity of the CKM matrix $V$. Although FCNCs appear within the Standard Model via one-loop and higher order diagrams with the help of the flavor-changing $W^{\pm}$ gauge bosons as internal lines,  they are strongly suppressed by the Glashow-Iliopoulos-Maiani (GIM) mechanism~\cite{Glashow:1970gm}. However, the geometrical interaction given by Eq.~(\ref{geom-action.2}) contributes to FCNCs at tree level due to the presence of the off-diagonal flavor terms since the matrices $\Lambda^Q_L$, $\Lambda^u_R$ and $\Lambda^d_R$ are not diagonal. Let us now briefly discuss some of the processes affected by FCNC-like tree-level diagrams arising fron the geometrical interaction.

We can write the interaction term as
\begin{align}\label{current-current}
\mathscr L^q_{4\psi}
=&
-\frac12
\left(
J_L^u + J_L^d + J_R^u + J_R^d
\right)^2\notag \\
=&-\frac12 \Big[(J_L^u)^2 + (J_L^d)^2 + (J_R^u)^2 + (J_R^d)^2 \notag \\
& \quad\quad 
+ 2J_L^uJ_L^d
+ 2J_L^uJ_R^u
+ 2J_L^uJ_R^d
+ 2J_L^dJ_R^u
+ 2J_L^dJ_R^d
+ 2J_R^uJ_R^d
\Big]\,,
\end{align}
where the up-type quark currents are given by 
\begin{align}
J_L^u &=\sum_{A,I,J}\lambda^Q_{AL}\;\bar u_{JL}\,(V_1^\dagger)_{JA}\,\gamma^a\,(V_1)_{AI}\,u_{IL},\\[4pt] J_R^u &=\sum_{A,I,J}\lambda^u_{AR}\;\bar u_{JR}\,(V_2^\dagger)_{JA}\,\gamma^a\,(V_2)_{AI}\,u_{IR},
\end{align}
and the down-type quark currents are given by 
\begin{align}
J_L^d &=\sum_{A,I,J}\lambda^Q_{AL}\;\bar d_{JL}\,(V^\dagger V_1^\dagger)_{JA}\,\gamma^a\,(V_1 V)_{AI}\,d_{IL},\\[4pt] J_R^d &=\sum_{A,I,J}\lambda^d_{AR}\;\bar d_{JR}\,(V_3^\dagger)_{JA}\,\gamma^a\,(V_3)_{AI}\,d_{IR}.
\end{align}
We note the appearance of the flavor structures
\begin{equation}
(\Lambda_L^u)_{JI}
=
\sum_A \lambda^Q_{AL}
\,(V_1^\dagger)_{JA}(V_1)_{AI}, \qquad (\Lambda_R^u)_{JI}
=
\sum_A \lambda^u_{AR}(V_2^\dagger)_{JA}(V_2)_{AI},
\end{equation}
\begin{equation}
(\Lambda_L^d)_{JI}
=
\sum_A \lambda^Q_{AL}
\,(V^\dagger V_1^\dagger)_{JA}(V_1 V)_{AI}, \qquad (\Lambda_R^d)_{JI}
=
\sum_A \lambda^d_{AR}(V_3^\dagger)_{JA}(V_3)_{AI}.
\end{equation}
Here, we have introduced $\Lambda_L^u$ which is identical to $\Lambda_L^Q$ of Eq.~\ref{coupling}, and $\Lambda_L^d=V^\dagger\Lambda_L^uV$. We can now write the currents in terms of the flavor structures as 
\begin{subequations}
\begin{equation}
J_L^u = \sum_{I,J}\bar u_{JL}\gamma^a(\Lambda_L^u)_{JI}u_{IL},    
\qquad J_R^u = \sum_{I,J}\bar u_{JR}\gamma^a(\Lambda_R^u)_{JI}u_{IR},    
\end{equation}
\begin{equation}
J_L^d = \sum_{I,J}\bar d_{JL}\gamma^a(\Lambda_L^d)_{JI}d_{IL},    
\qquad
J_R^d = \sum_{I,J}\bar d_{JR}\gamma^a(\Lambda_R^d)_{JI}d_{IR}.    
\end{equation}
\end{subequations}
These are the expressions for the currents to be used in~\eqref{current-current}. Since the interaction involves only quarks, it contributes exclusively to
non-leptonic processes including $\Delta F = 1$ (non-leptonic weak transitions) and $\Delta F = 2$ (meson mixing) processes.



\subsection{$\Delta F = 1$ Non-Leptonic Transitions}

The four-quark interaction gives rise to $\Delta F = 1$ non-leptonic flavor-changing neutral current (FCNC) transitions at tree level once the quark fields are rotated to the mass basis. In particular, the flavor structures $\Lambda_L^d$, $\Lambda_R^d$, $\Lambda_L^u$, and $\Lambda_R^u$ introduced above are, in general, non-diagonal in the quark mass basis. Consequently, the current--current interaction in Eq.~(\ref{current-current}) contains terms involving off-diagonal quark bilinears such as $\bar d_J\gamma^a d_I$ and $\bar u_J\gamma^a u_I$ with $I\neq J$. When these currents are multiplied together, they generate four-quark operators of the form $(\bar d_J\gamma^a d_I)(\bar u_K\gamma_a u_L)$ with $I\neq J$ and $K=L$ that mediate $\Delta F = 1$ transitions. These operators mediate transitions which are flavor-changing but neutral in charge. Unlike in the Standard Model, where FCNC processes arise only at loop level via penguin and box diagrams and are therefore suppressed by the Glashow--Iliopoulos--Maiani mechanism, the present interaction generates such transitions at tree level, potentially leading to sizable deviations from the Standard Model. In hadronic processes, these operators contribute to a wide class of non-leptonic decay modes, including $K \to \pi\pi$, $B \to K\pi$, $B \to \phi K$, $B \to \pi\pi$, and charm decays such as $D \to \pi\pi$ and $D \to K\bar K$. The underlying quark-level processes are $d \to s \bar{q} q$, $b \to s \bar{q} q$, $b \to d \bar{q} q$, and $c \to u \bar{q} q$. We will discuss the first process in detail below and the operators relevant to the other processes can be found in Appendix~\ref{operators}.



Let us now look at a specific channel of a $\Delta F=1$ decay in detail, specifically, $K^+ \to \pi^0 \pi^+$. We can write the effective Hamiltonian from the geometrical interaction relevant for this decay as 
\begin{equation}
\mathcal H_{\text{eff}}^{K^+ \to \pi^0 \pi^+}
=
\sum_{q=u,d}
\Big[
C_{q,LL}^{(1)} \mathcal O_{q,LL}^{(1)}
+
C_{q,LR}^{(1)} \mathcal O_{q,LR}^{(1)}
+
C_{q,RL}^{(1)} \mathcal O_{q,RL}^{(1)}
+
C_{q,RR}^{(1)} \mathcal O_{q,RR}^{(1)}
\Big],
\end{equation}
where
\begin{align}
\mathcal O_{q,LL}^{(1)} &= (\bar s_L \gamma^a d_L)(\bar q_L \gamma_a q_L), \qquad \mathcal O_{q,LR}^{(1)} = (\bar s_L \gamma^a d_L)(\bar q_R \gamma_a q_R), \\
\mathcal O_{q,RL}^{(1)} &= (\bar s_R \gamma^a d_R)(\bar q_L \gamma_a q_L), \qquad \mathcal O_{q,RR}^{(1)} = (\bar s_R \gamma^a d_R)(\bar q_R \gamma_a q_R),
\end{align} 
and we have defined the ``geometrical Wilson coefficients'' $C_q$ as 
\begin{align}C_{u,LL}^{(1)} &= (\Lambda_L^d)_{21}\,(\Lambda_L^u)_{11}, \qquad C_{d,LL}^{(1)} = \frac{1}{2}(\Lambda_L^d)_{21}\,(\Lambda_L^d)_{11},\\
C_{q,LR}^{(1)} &= (\Lambda_L^d)_{21}\,(\Lambda_R^q)_{11}, \qquad C_{q,RL}^{(1)} = (\Lambda_R^d)_{21}\,(\Lambda_L^q)_{11},\\
C_{u,RR}^{(1)} &= (\Lambda_R^d)_{21}\,(\Lambda_R^u)_{11}, \qquad C_{d,RR}^{(1)} = \frac{1}{2}(\Lambda_R^d)_{21}\,(\Lambda_R^d)_{11}.\end{align} 
The operators $\mathcal O_{q,RL}^{(1)}$ and $\mathcal O_{q,RR}^{(1)}$ are present in the geometrical interaction but not in the SM, whereas the operator $(\bar s \gamma^a P_L u)(\bar u \gamma_a P_L d)$, which appears in the SM, is not present here. The SM operators for this process can be found in~\cite{Buras:1991jm,Buras:1993dy,Ciuchini:1993vr,Buchalla:1995vs}. 

Let us also write down the expressions for these coefficients in terms of the flavor-mixing matrices $V, V_1, V_2, V_3$ and the couplings $\lambda^Q_L,\lambda^u_R,\lambda^d_R$,
\begin{align}
C_{u,LL}^{(1)}
=&
\left[
\sum_A \lambda^Q_{AL}(V^\dagger V_1^\dagger)_{2A}(V_1 V)_{A1}
\right]
\left[
\sum_B \lambda^Q_{BL}(V_1^\dagger)_{1B}(V_1)_{B1}
\right]\,,  \notag \\
C_{d,LL}^{(1)}
=&
\frac{1}{2}\left[
\sum_A \lambda^Q_{AL}(V^\dagger V_1^\dagger)_{2A}(V_1 V)_{A1}
\right]
\left[
\sum_B \lambda^Q_{BL}(V^\dagger V_1^\dagger)_{1B}(V_1 V)_{B1}
\right]\,, \notag \\
C_{u,LR}^{(1)}
=&
\left[
\sum_A \lambda^Q_{AL}(V^\dagger V_1^\dagger)_{2A}(V_1 V)_{A1}
\right]
\left[
\sum_B \lambda^u_{BR}(V_2^\dagger)_{1B}(V_2)_{B1}
\right]\,, \notag \\
C_{d,LR}^{(1)}
=&
\left[
\sum_A \lambda^Q_{AL}(V^\dagger V_1^\dagger)_{2A}(V_1 V)_{A1}
\right]
\left[
\sum_B \lambda^d_{BR}(V_3^\dagger)_{1B}(V_3)_{B1}
\right]\,, \notag \\
C_{u,RL}^{(1)}
=&
\left[
\sum_A \lambda^d_{AR}(V_3^\dagger)_{2A}(V_3)_{A1}
\right]
\left[
\sum_B \lambda^Q_{BL}(V_1^\dagger)_{1B}(V_1)_{B1}
\right]\,, \notag \\
C_{d,RL}^{(1)}
=&
\left[
\sum_A \lambda^d_{AR}(V_3^\dagger)_{2A}(V_3)_{A1}
\right]
\left[
\sum_B \lambda^Q_{BL}(V^\dagger V_1^\dagger)_{1B}(V_1 V)_{B1}
\right]\,, \notag \\
C_{u,RR}^{(1)}
=&
\left[
\sum_A \lambda^d_{AR}(V_3^\dagger)_{2A}(V_3)_{A1}
\right]
\left[
\sum_B \lambda^u_{BR}(V_2^\dagger)_{1B}(V_2)_{B1}
\right]\,, \notag \\
C_{d,RR}^{(1)}
=&
\frac{1}{2}\left[
\sum_A \lambda^d_{AR}(V_3^\dagger)_{2A}(V_3)_{A1}
\right]
\left[
\sum_B \lambda^d_{BR}(V_3^\dagger)_{1B}(V_3)_{B1}
\right]\,.
\end{align}
The final state $\ket{\pi^0 \pi^+}$ has total isospin $I=2$, and therefore it is a $\Delta I=\frac{3}{2}$ transition. Let us rewrite the Hamiltonian, suppressing the chirality indices, as
\begin{equation}
\mathcal H_{\text{eff}}
=\frac{C^{(1)}_u + C^{(1)}_d}{2}(\bar s d)(\bar u u + \bar d d)
+
\frac{C^{(1)}_u - C^{(1)}_d}{2}(\bar s d)(\bar u u - \bar d d).
\end{equation}
Since the operator $(\bar s d)(\bar u u + \bar d d)$ induces only $\Delta I=\frac{1}{2}$ transitions, its matrix element vanishes:
%
%
\begin{equation}
\langle \pi^0 \pi^+ | (\bar s d)(\bar u u + \bar d d) | K^+ \rangle = 0.
\end{equation}
%
%
%
Therefore, suppressing the chirality indices, we can write the decay amplitude due to the geometrical interaction as
\begin{equation}
\mathcal A_{G}(K^+ \to \pi^0 \pi^+)
=\frac{C^{(1)}_u - C^{(1)}_d}{2}
\langle \pi^0 \pi^+ | (\bar s d)(\bar u u - \bar d d) | K^+ \rangle.
\end{equation}
As discussed earlier, calculations become much simpler in the special case when the torsion basis is aligned with the flavor basis. We will therefore calculate the amplitude assuming $V_i=\text{diag}(e^{i\zeta^{(i)}_1},e^{i\zeta^{(i)}_2},e^{i\zeta^{(i)}_3})$. In this case, only the coefficients $C^{(1)}_{q,LL}$ and $C^{(1)}_{q,LR}$ corresponding to the operators $(\bar s_L \gamma^a d_L)(\bar q_L \gamma_a q_L)$ and $(\bar s_L \gamma^a d_L)(\bar q_R \gamma_a q_R)$ remain, while $C^{(1)}_{q,RL}$ and $C^{(1)}_{q,RR}$ become 0. Therefore, we find 
\begin{align}
C^{(1)}_{u,LL} & =(\Lambda_L^d)_{21}(\Lambda_L^u)_{11}, \qquad C^{(1)}_{d,LL} =\frac{1}{2}(\Lambda_L^d)_{21}(\Lambda_L^d)_{11}, \notag \\
C^{(1)}_{u,LR} & =(\Lambda_L^d)_{21}(\Lambda_R^u)_{11}, \qquad C^{(1)}_{d,LR}  =(\Lambda_L^d)_{21}(\Lambda_R^d)_{11}.   
\end{align}
The flavor structures are simplified in this case, 
\begin{align}
(\Lambda_L^d)_{21}&=\sum_A \lambda^Q_{AL}(V^\dagger)_{2A}(V)_{A1}\notag\\&=\lambda^Q_{1L} V^*_{us}V_{ud}+\lambda^Q_{2L} V^*_{cs}V_{cd}+\lambda^Q_{3L} V^*_{ts}V_{td}\notag\\&=\lambda^Q_{1L} s_{12}c_{12}c_{13}^2+\lambda^Q_{2L}\Big(-c_{12}s_{12}c_{23}^2-c_{12}^2 c_{23}s_{23}s_{13} e^{i\delta}+s_{12}^2 s_{23}c_{23}s_{13} e^{-i\delta}+s_{12}c_{12}s_{23}^2 s_{13}^2\Big)\notag\\&\qquad\qquad+\lambda^Q_{3L}\Big(-c_{12}s_{12}s_{23}^2+c_{12}^2 s_{23}c_{23}s_{13} e^{i\delta}-s_{12}^2 c_{23}s_{23}s_{13} e^{-i\delta}+s_{12}c_{12}c_{23}^2 s_{13}^2\Big), \\
(\Lambda_L^d)_{11}&=\sum_A \lambda^Q_{AL}(V^\dagger)_{1A}(V)_{A1}\notag\\&=\lambda^Q_{1L}|V_{ud}|^2+
\lambda^Q_{2L}|V_{cd}|^2+\lambda^Q_{3L}|V_{td}|^2\notag\\&=\lambda^Q_{1L}c_{12}^2 c_{13}^2+\lambda^Q_{2L}\Big(s_{12}^2 c_{23}^2+c_{12}^2 s_{23}^2 s_{13}^2+2 s_{12}c_{12}c_{23}s_{23}s_{13}\cos\delta\Big)\notag\\
&\qquad\qquad+\lambda^Q_{3L}\Big(s_{12}^2 s_{23}^2+c_{12}^2 c_{23}^2 s_{13}^2-2 s_{12}c_{12}s_{23}c_{23}s_{13}\cos\delta\Big),
\end{align}
\begin{align}
(\Lambda_L^u)_{11}=\lambda^Q_{1L}, \qquad (\Lambda_R^u)_{11}=\lambda^u_{1R}, \qquad (\Lambda_R^d)_{11}=\lambda^d_{1R}.\qquad\qquad\qquad\qquad\qquad\qquad\qquad
\end{align} 
If the $\lambda^Q$ couplings are universal, that is, $\lambda^Q_{1L}=\lambda^Q_{2L}=\lambda^Q_{3L}$, then the quantity $(\Lambda_L^d)_{21}=0$ due to the unitarity of the CKM matrix, and all remaining geometrical Wilson coefficients involving the $\lambda$'s vanish. Thus a universal geometrical interaction will not contribute to this decay.

The decay amplitude due to the geometrical interaction is therefore
\begin{equation}
\mathcal A_{G}(K^+ \to \pi^0 \pi^+)
=\frac{1}{2}\,(C^{(1)}_{u,LL} - C^{(1)}_{d,LL})\mathcal M_1+\frac{1}{2}\,(C^{(1)}_{u,LR} - C^{(1)}_{d,LR})\mathcal M_2,
\end{equation}
 where the hadronic matrix elements $\mathcal M_1$ and $\mathcal M_2$ are given by
\begin{equation}
\mathcal M_1 
=\langle \pi^0 \pi^+ | (\bar s_L \gamma^a d_L)(\bar u_L \gamma_a u_L-\bar d_L \gamma_a d_L) | K^+ \rangle,
\end{equation}
\begin{equation}
\mathcal M_2 
=\langle \pi^0 \pi^+ | (\bar s_L \gamma^a d_L)(\bar u_R \gamma_a u_R-\bar d_R \gamma_a d_R) | K^+ \rangle.
\end{equation}
Matrix elements for nonleptonic decays are in general much more difficult to calculate than those for semi-leptonic and leptonic decays. This is because a naive factorization of matrix elements of four-quark operators into a product of matrix elements of two quark bilinears is not very reliable. However, we can still use the vacuum insertion approximation (VIA) to estimate $\mathcal M_1$ and $\mathcal M_2$. This approximation is commonly used in the phenomenological treatment of non-leptonic kaon decays~\cite{Buras:1993dy, Bigi:2000yz, Donoghue:2022wrw, Carrasco:2013jda}. In the VIA, we factorize the matrix elements by inserting the vacuum as an intermediate state in all possible ways. Thus, we can write the matrix element
\begin{align}
\mathcal M_1 &= 
\langle \pi^0\,\pi^+\,|\,(\bar s\,\gamma^\mu P_L\, d)\,(\bar u\,\gamma_\mu P_L\,u - \bar d\,\gamma_\mu P_L\,d)\,|\,K^+\rangle\\
&\approx
\langle \pi^0 |\,\bar u\gamma^\mu P_L u - \bar d\gamma^\mu P_L d\,|0\rangle
\langle \pi^+ |\,\bar s\gamma_\mu P_L d\,|K^+\rangle.\label{factor}
\end{align}
Let us take the the four-momenta of the mesons as $p_K$, $p_1$ and $p_2$ for $K^+$, $\pi^0$ and $\pi^+$ respectively. For the vacuum-to-pion matrix element, only the axial current contributes and we have
\begin{equation}
\langle \pi^0 | \bar u\gamma^\mu\gamma_5 u |0\rangle = -\langle \pi^0 | \bar d\gamma^\mu\gamma_5 d |0\rangle = \frac{i f_\pi}{\sqrt{2}}\, p_1^\mu.\end{equation}
Therefore, we get
\begin{equation}
\langle \pi^0 | \bar u\gamma^\mu P_L u - \bar d\gamma^\mu P_L d |0\rangle
=
-\frac{i}{\sqrt{2}} f_\pi\, p_1^\mu.
\end{equation}
The matrix element of the vector current in the other factor can be parameterized as~\cite{Becirevic:2004ya,Belyaev:2000xt}
\begin{equation}
\langle \pi^+ | \bar s \gamma^\mu d | K^+\rangle
=
(p_K + p_2)^\mu f_+(q^2)
+
(p_K - p_2)^\mu f_-(q^2),
\end{equation}
while the axial current vanishes. Here, $q=p_K-p_2$. Thus,\begin{equation}
\mathcal M_1=
-\frac{i}{2\sqrt{2}} f_\pi
\left[
p_1 \cdot (p_K + p_2)\, f_+
+
p_1 \cdot (p_K - p_2)\, f_-
\right].
\end{equation}

Since $p_K = p_1 + p_2$, we can calculate $p_1 \cdot p_2 = \frac{1}{2}(m_K^2 - 2 m_\pi^2)$, $p_1 \cdot (p_K + p_2) =m_K^2 - m_\pi^2$, and $p_1 \cdot (p_K - p_2) = m_\pi^2$. Hence, \begin{equation}
\mathcal M_1
=
-\frac{i}{2\sqrt{2}} f_\pi
\left[
(m_K^2 - m_\pi^2)\, f_+(q^2)
+
m_\pi^2\, f_-(q^2)
\right].
\end{equation} 
For the other matrix element $\mathcal{M}_2$, we can again follow the prescription of using the VIA. Then we will get a factorization like Eq.~(\ref{factor}), with the right-chiral projection $P_R$ instead of $P_L$ in the vacuum-to-pion factor, while the other factor remains unchanged. Thus, using the VIA , we get the matrix element,
\begin{equation}
\mathcal M_2
=\frac{i}{2\sqrt{2}} f_\pi
\left[
(m_K^2 - m_\pi^2)\, f_+(q^2)
+
m_\pi^2\, f_-(q^2)
\right].
\end{equation}
Therefore, we can write the amplitude due to the geometrical interaction as
\begin{equation}
\mathcal A_{G}(K^+ \to \pi^0 \pi^+)
=\frac{i}{4\sqrt{2}}(C^{(1)}_{d,LL} - C^{(1)}_{u,LL} + C^{(1)}_{u,LR} - C^{(1)}_{d,LR})\, f_\pi
\left[
(m_K^2 - m_\pi^2) f_+
+
m_\pi^2 f_-
\right].
\end{equation}

This has to be added to the amplitude due to the SM interactions. The above expressions were obtained under the simplifying assumption that the torsion basis is aligned with the flavor basis. In general however, all eight coefficients $C^{(1)}_{q,LL}$, $C^{(1)}_{q,LR}$, $C^{(1)}_{q,RL}$ and $C^{(1)}_{q,RR}$ are non-vanishing.
Non-alignment introduces three additional unitary matrices $V_1$, $V_2$, and $V_3$, which describe the mismatch between the torsion and the flavor basis of the left-handed quark doublets, right-handed up-type quarks and right-handed down-type quarks, respectively.
Consequently, the geometrical Wilson coefficients depend not only on the geometrical couplings $\lambda^Q_L$, $\lambda^u_R$, and $\lambda^d_R$, but also on the additional mixing parameters contained in $V_1$, $V_2$, and $V_3$. Unlike the aligned case, the right-handed flavor-changing coupling $(\Lambda_R^d)_{21}$ is generally non-zero. As a result, when the torsion basis is not aligned with the flavor basis, the operators $(\bar s_R\gamma^\mu d_R)(\bar q_L\gamma_\mu q_L)$ and $(\bar s_R\gamma^\mu d_R)(\bar q_R\gamma_\mu q_R)$, represented by the coefficients $C^{(1)}_{q,RL}$ and $C^{(1)}_{q,RR}$, contribute to the decay amplitude. These operators, which cannot be ruled out using theoretical arguments, have no counterpart in the Standard Model and therefore constitute a distinctive signature of the geometrical interaction. 

There are qualitative differences between the geometrical interaction and the SM contribution to the decay $K^+\to\pi^0\pi^+$. The SM decay is dominated by the current-current operator $(\bar u\gamma^a P_L d)(\bar s\gamma_a P_L u)$, although penguin operators are also present as subleading contributions~\cite{Buras:2014maa,Buras:2016fys,Carrasco:2013jda}. In contrast, the geometrical interaction generates four-quark operators of the form $(\bar s\gamma^a P_{L,R}d)(\bar q\gamma_a P_{L,R}q)$ of which only $(\bar s\gamma^a P_{L}d)(\bar q\gamma_a P_{L,R}q)$ contribute in the case of alignment. These operators also appear in the SM as QCD penguin operators, but they do not contribute in the isospin-symmetric limit. As mentioned earlier, in the absence of alignment, the non-SM operators $(\bar s\gamma^a P_{R}d)(\bar q\gamma_a P_{L,R}q)$ also contribute to the decay.

The analysis presented above for the decay
$K^+\rightarrow\pi^0\pi^+$ can be extended straightforwardly to other
non-leptonic kaon decay channels. The effective Hamiltonian retains the same
operator structure, and the corresponding geometrical Wilson coefficients are obtained by
selecting the appropriate flavor-changing entries of the matrices
$\Lambda_{L,R}^{u,d}$. The principal differences arise from the flavor content
of the external mesons and the isospin of the final state.

For the charged decay $K^-\rightarrow\pi^0\pi^-$, the effective Hamiltonian is
simply the Hermitian conjugate of that for $K^+\rightarrow\pi^0\pi^+$. The
coefficients $C_q$ are therefore replaced by their complex conjugates, while
the hadronic matrix elements are related by CP symmetry up to possible
CP-violating phases. Consequently, in the absence of CP violation, the decay
rates of $K^+$ and $K^-$ are identical, whereas any difference between the two
provides a direct probe of the CP-violating phases contained in the mixing 
matrices $V$, $V_1$, $V_2$, and $V_3$.

For neutral kaon decays, such as
$K^0(\bar sd)\rightarrow\pi^+\pi^-$ and
$K^0(\bar sd)\rightarrow\pi^0\pi^0$, the flavor-changing transition remains
$s\rightarrow d$, so the same off-diagonal flavor structures,
$(\Lambda_L^d)_{21}$ and, in the non-aligned case,
also $(\Lambda_R^d)_{21}$, govern the dynamics. However, unlike the
$K^+\rightarrow\pi^0\pi^+$ channel, the final two-pion state contains both
isospin $I=0$ and $I=2$ components. Consequently, both the operator
$(\bar sd)(\bar uu+\bar dd)$, which induces $\Delta I=\frac12$ transitions,
and the operator
$(\bar sd)(\bar uu-\bar dd)$, contribute simultaneously. The neutral kaon
channels therefore probe both combinations
$C_u^{(1)}+C_d^{(1)}$ and
$C_u^{(1)}-C_d^{(1)}$, whereas the
$K^+\rightarrow\pi^0\pi^+$ decay isolates only the
$\Delta I=\frac32$ contribution proportional to
$C_u^{(1)}-C_d^{(1)}$. As a result, neutral kaon decays are considerably more
sensitive to the flavor structure of the geometrical interaction and provide a
natural laboratory for studying both the $\Delta I=\frac12$ enhancement and
CP-violating observables~\cite{Cirigliano:2003gt,RBC:2020kdj}.

So far we have only discussed $\Delta F=1$ transitions and how these are affected due to the geometrical interaction. The geometrical interaction also contributes to $\Delta F=2$ neutral meson mixings like $K^0$--$\bar{K}^0$, $B_d^0$--$\bar{B}_d^0$, $B_s^0$--$\bar{B}_s^0$, and $D^0$--$\bar{D}^0$. We will study them in a future work.


\section{Rephasing Invariants from the Geometrical Interaction}\label{Jarlskog}
The geometrical interaction introduced in Eq.~(\ref{geom-action}) contains three new Hermitian flavor matrices, $\Lambda_L^Q,\Lambda_R^u,\Lambda_R^d$, in addition to the usual unitary CKM matrix $V$. Consequently, the Standard Model's single Jarlskog invariant is no longer sufficient to characterize CP violation. The physical CP-violating quantities are those combinations of the flavor matrices whose imaginary part changes under arbitrary rephasings of the quark fields.

In the Standard Model, there is only a single physical CP-violating phase, and all CP-odd observables are proportional to the Jarlskog invariant~\cite{Jarlskog:1985cw,Jarlskog:1985ht}
\begin{equation}
J_{\rm CKM}
=
{\rm Im}
\left(
V_{AB}V_{CD}V^*_{AD}V^*_{CB}
\right),
\end{equation}
for any choice of distinct indices $A \neq C$ and $B \neq D$. A commonly used choice is
\begin{equation}
J_{\rm CKM}
=
{\rm Im}
\left(
V_{ud}V_{cs}V_{us}^*V_{cd}^*
\right).
\end{equation}
All CP-violating observables in the Standard Model vanish if 
\begin{equation}
J_{\rm CKM}=0.
\end{equation}
The quark mass eigenstates may be rephased according to
\begin{equation}\label{rephasing}
\tilde{u}_A = e^{i\phi_A}u_A,\qquad
\tilde{d}_A = e^{i\chi_A}d_A,
\end{equation}
under which the CKM matrix transforms as
\begin{equation}\label{CKM-new}
\tilde{V}_{AB} = e^{i(\chi_B-\phi_A)}V_{AB},
\end{equation}
and the flavor matrices $\Lambda_L^Q,\Lambda_R^u,\Lambda_R^d$ as
\begin{align}
(\tilde{\Lambda}_L^Q)_{AB}& = e^{i(\phi_B-\phi_A)}({\Lambda}_L^Q)_{AB},\label{flav1-new}\\
(\tilde{\Lambda}_R^u)_{AB}& = e^{i(\phi_B-\phi_A)}({\Lambda}_R^u)_{AB},\label{flav2-new}\\
(\tilde{\Lambda}_R^d)_{AB}& = e^{i(\chi_B-\chi_A)}({\Lambda}_R^d)_{AB}\label{flav3-new}.
\end{align}
There is no sum in the $A,B$ indices in Eqs.~(\ref{rephasing},\ref{CKM-new},\ref{flav1-new},\ref{flav2-new},\ref{flav3-new}). Rephasing invariants in the SM are the moduli of the CKM matrix elements and quartets of the form $(V_{AB}V_{CD}V^*_{AD}V^*_{CB})$ with $A \neq C$ and $B \neq D$ and all other rephasing invariants in the SM can be constructed using these~\cite{Branco:1987mj,Branco:1999fs,Wu:1985ea}.

In our case, in the presence of the geometrical interaction, the simplest rephasing invariants are the diagonal elements and the moduli of the off-diagonal elements of the matrices $\Lambda_L^Q,\Lambda_R^u,\Lambda_R^d$. Let us define $X\equiv \Lambda^Q_L$, $Y\equiv \Lambda^u_R$ and $Z\equiv \Lambda^d_R$ for notational convenience. Therfore, for $\Lambda_L^Q$, we have the independent rephasing invariants $X_{11}$, $X_{22}$, $X_{33}$, $\abs{X_{12}}^2$, $\abs{X_{23}}^2$, $\abs{X_{13}}^2$. Similarly, for $\Lambda_R^u$ and $\Lambda_R^d$, we have 12 more rephasing invariants as the ones above but with $X$ replaced by $Y$ and $Z$, respectively.  

The simplest mixed invariant is
\begin{equation}
R^{AB}=
(X)_{AB}
(Y)^*_{AB}.    
\end{equation}
The lowest order invariants involving a combination of the CKM matrix and the other flavor matrices consists of elements of six mixing matrices given by
\begin{equation}\label{mixed}
S_1^{ABCD}=V_{AB}X_{CA}V^*_{CD}Z^*_{DB}, \qquad 
S_2^{ABCD}=V_{AB}Y_{CA}V^*_{CD}Z^*_{DB}.   
\end{equation}
Let us next look at cyclic rephasing invariants of the form 
\begin{align}
C^{A_1A_2...A_n}_L
&=X_{A_1A_2}
X_{A_2A_3}...
X_{A_nA_1}\notag\\
&=(D^{(1)\dagger}\tilde V_1^\dagger D_\lambda\tilde V_1D^{(1)})_{A_1A_2}(D^{(1)\dagger}\tilde V_1^\dagger D_\lambda\tilde V_1
D^{(1)})_{A_2A_3}...(D^{(1)\dagger}\tilde V_1^\dagger D_\lambda\tilde V_1D^{(1)})_{A_nA_1}\notag\\
&=(\tilde V_1^\dagger D_\lambda\tilde V_1)_{A_1A_2}
(\tilde V_1^\dagger D_\lambda\tilde V_1)_{A_2A_3}...(\tilde V_1^\dagger D_\lambda\tilde V_1)_{A_nA_1}\notag\\
&=(X^{(1)})_{A_1A_2}
(X^{(1)})_{A_2A_3}...
(X^{(1)})_{A_nA_1},
\end{align}
where $D_\lambda=\mathrm{diag}(\lambda^Q_{1L},\lambda^Q_{2L},\lambda^Q_{3L})$ and $X^{(1)}=\tilde V_1^\dagger D_\lambda\tilde V_1$. We see that the phases $\alpha^{(1)}_1$ and $\alpha^{(1)}_2$ cancel out in the product. Similarly, for the invariants
\begin{equation}
C^{A_1A_2...A_n}_{u}
=Y_{A_1A_2}
Y_{A_2A_3}...
Y_{A_nA_1}, \qquad C^{A_1A_2...A_n}_{d}
=Z_{A_1A_2}
Z_{A_2A_3}...
Z_{A_nA_1},   
\end{equation}
the phases $\alpha^{(2)}_1$, $\alpha^{(2)}_2$ and $\alpha^{(3)}_1$, $\alpha^{(3)}_2$ drop out. The invariant $C^{A_1A_2...A_n}_L$ therefore consists of seven free parameters, four from the matrix $\tilde V_1$ and the three couplings $\lambda^Q_{1L},\lambda^Q_{2L},\lambda^Q_{3L}$. Similar conclusions hold for the invariants $C^{A_1A_2...A_n}_{u}$ and $C^{A_1A_2...A_n}_{d}$. We already have six independent rephasing invariants from the diagonal elements and the moduli of the off-diagonal elements. The invariant ${\rm Im}(C^{123}_L)$, which we will talk about later, along with the six above therefore constitute seven independent rephasing invariants. Since all other higher order cyclic invariants $C^{A_1A_2...A_n}_L$, $n>3$ are dependent on the seven parameters, they can be constructed out of $X_{11}$, $X_{22}$, $X_{33}$, $\abs{X_{12}}^2$, $\abs{X_{23}}^2$, $\abs{X_{13}}^2$ and ${\rm Im}(C^{123}_L)$.   

For example at $n=4$, the cyclic invariants consisting of elements of eight mixing matrices are
\begin{align}
C_L^{ABCD}
&=
X_{AB}
X_{BC}
X_{CD}
X_{DA},\\
C_{u}^{ABCD}
&=
Y_{AB}
Y_{BC}
Y_{CD}
Y_{DA},\\
C_{d}^{ABCD}
&=
Z_{AB}
Z_{BC}
Z_{CD}
Z_{DA}.
\end{align}
These are also the direct analogues of the invariant quartets $(V_{AB}V_{CD}V^*_{AD}V^*_{CB})$ for $A\neq C$ and $B\neq D$.

Similarly we can construct mixed cyclic invariants consisting of elements of six mixing matrices, 
\begin{equation}
M_1^{ABC}= 
X_{AB}
Y_{BC}
X_{CA}, \qquad  M_2^{ABC}= 
Y_{AB}
X_{BC}
Y_{CA}.  
\end{equation}
In general for $n=$ odd, we have the mixed invariants 
\begin{equation}
M_1^{A_1A_2...A_n}= 
X_{A_1A_2}
Y_{A_2A_3}...
X_{A_nA_1}, \qquad  M_2^{A_1A_2...A_n}= 
Y_{A_1A_2}
X_{A_2A_3}...
Y_{A_nA_1},  
\end{equation}
and for $n=$ even,
\begin{equation}
M^{A_1A_2...A_n}= 
X_{A_1A_2}
Y_{A_2A_3}...
X_{A_{n-1}A_n}
Y_{A_nA_1}.  
\end{equation}
Analogous to the previous case, higher order mixed invariants involving eight or more mixing matrices should be expressible in terms of the lower order ones. However we have not explicitly worked out their relations.

CP is conserved if and only if all CP-odd rephasing invariants vanish. Therefore, the CP-violating effects arising from the geometrical interaction are encoded in the invariants like the ones given above involving the three new Hermitian flavor matrices.

Among these, let us try to work out the CP-odd part of the invariant $C^{123}_L$.
%
%
After some simplifications, the invariant turns out to be 
\begin{equation}
{\rm Im}(C^{123}_L)=-(\lambda^Q_{1L}-\lambda^Q_{2L})(\lambda^Q_{2L}-\lambda^Q_{3L})(\lambda^Q_{3L}-\lambda^Q_{1L}) \sin{a_1}\sin{b_1}\sin{c_1}\cos{a_1}\cos{c_1}\cos^2{b_1}sin{\delta_1}.  
\end{equation}
This expression has several important implications. First, the invariant is proportional to $\sin\delta_1$, and therefore the cyclic CP-odd invariant $\rm Im(C_L^{123})$ vanishes for $\delta_1=0,\pi$. More generally, when the geometrical couplings $\lambda^Q_L$ are non-degenerate and the mixing angles $a_1, b_1, c_1$ are non-zero, a non-vanishing $\sin\delta_1$ gives a non-zero value of this particular CP-odd invariant. The factor $$\sin a_1\sin b_1\sin c_1\cos a_1\cos c_1\cos^2 b_1\sin\delta_1$$ has the same structure as the Jarlskog invariant. In particular, $\rm Im(C_L^{123})$ vanishes if any one of the three mixing angles is zero, or if the complex phase $\delta_1=0,\pi$. The appearance of the Vandermonde factor $(\lambda^Q_{1L}-\lambda^Q_{2L})(\lambda^Q_{2L}-\lambda^Q_{3L})(\lambda^Q_{3L}-\lambda^Q_{1L})$ also shows that this CP-odd invariant vanishes whenever any two of the geometrical couplings $\lambda_L^Q$ are degenerate. This is analogous to the role played by eigenvalue differences in the Jarlskog invariant for CP violation in the Standard Model. The analogous cyclic invariants which can be constructed from $\Lambda_R^u$ and $\Lambda_R^d$ are 
\begin{equation}
{\rm Im}(C^{123}_{u})=-(\lambda^u_{1R}-\lambda^u_{2R})(\lambda^u_{2R}-\lambda^u_{3R})(\lambda^u_{3R}-\lambda^u_{1R}) \sin{a_2}\sin{b_2}\sin{c_2}\cos{a_2}\cos{c_2}\cos^2{b_2}sin{\delta_2},  
\end{equation}
\begin{equation}
{\rm Im}(C^{123}_{d})=-(\lambda^d_{1R}-\lambda^d_{2R})(\lambda^d_{2R}-\lambda^d_{3R})(\lambda^d_{3R}-\lambda^d_{1R}) \sin{a_3}\sin{b_3}\sin{c_3}\cos{a_3}\cos{c_3}\cos^2{b_3}sin{\delta_3}. 
\end{equation}

We should note that all the information about CP violation due to the geometrical interaction is not encoded in the cyclic invariants. Although the phases $\alpha^{(1)}_1$, $\alpha^{(1)}_2$ have disappeared from the expression of $\rm Im(C_L^{123})$, it does not imply that these phases are unphysical in the full theory. We will see now that these and similar phases will show up in the mixed rephasing invariants such as $S_1^{ABCD}$ and $S_2^{ABCD}$. The cyclic invariants encode the CP-odd information associated with the torsional coupling matrices $X,Y,Z$, whereas the mixed invariants characterize CP violation arising from the relative orientation and relative phases of these matrices.

Let us now work out the invariants $S_1^{ABCD}$ and $S_2^{ABCD}$. Using Eq.~\ref{mixed}, we find that the phases $\alpha^{(1)}_1$ and $\alpha^{(1)}_2$ appear in the invariants ${\rm Im}(S_1^{1121})$ and ${\rm Im}(S_1^{1131})$, respectively. The full expression for ${\rm Im}(S_1^{1121})$ is given in Appendix~\ref{invariants}. Similarly, the invariants ${\rm Im}(S_1^{1112})$ and ${\rm Im}(S_1^{1113})$ depend on the phases $\alpha^{(3)}_1$ and $\alpha^{(3)}_2$, respectively, while ${\rm Im}(S_2^{1211})$ and ${\rm Im}(S_2^{1311})$ depend on $\alpha^{(2)}_1$ and $\alpha^{(2)}_2$, respectively.

We have already found 18 independent rephasing invariants due to the diagonal and off-diagonl entries in $X,Y,Z$. Now, we have found three more cyclic invariants ${\rm Im}(C^{123}_{L})$, ${\rm Im}(C^{123}_{u})$, ${\rm Im}(C^{123}_{d})$ and six mixed invariants ${\rm Im}(S_1^{1121})$, ${\rm Im}(S_1^{1131})$, ${\rm Im}(S_2^{1211})$, ${\rm Im}(S_2^{1311})$, ${\rm Im}(S_1^{1112})$, and ${\rm Im}(S_1^{1113})$. Therefore we have constructed 27 independent rephasing invariants for the 27 new free parameters due to the geometrical interaction. 

In the case of alignment, all of these 27 invariants vanish, only the Jarlskog invariant remains. However, the Jarlskog invariant itself is modified due to the geometrical interaction. We discuss this in the next section.

\section{Jarlskog Invariant Revisited}\label{Revisited}

The geometrical interaction of Eq.~(\ref{current-current}) can be rewritten in a form that exhibits up-down mixing by performing a Fierz rearrangement. The relevant Fierz identities for left-handed, right-handed and left-right vector currents is respectively
\begin{equation}
(\bar\psi_L\gamma^a\chi_L)
(\bar\eta_L\gamma_a\xi_L)
=
(\bar\psi_L\gamma^a\xi_L)
(\bar\eta_L\gamma_a\chi_L).
\label{LLFierz}
\end{equation}
\begin{equation}
(\bar\psi_R\gamma^a\chi_R)
(\bar\eta_R\gamma_a\xi_R)
=
(\bar\psi_R\gamma^a\xi_R)
(\bar\eta_R\gamma_a\chi_R).
\end{equation}
\begin{equation}
(\bar\psi_L\gamma^a\chi_L)
(\bar\eta_R\gamma_a\xi_R)
=
-2
(\bar\psi_L\xi_R)
(\bar\eta_R\chi_L).
\label{LRFierz}
\end{equation}
After the Fierz rearrangement, we are left with six terms involving purely up or down-type quarks from $(J^u_L)^2$, $(J^d_L)^2$, $(J^u_R)^2$, $(J^d_R)^2$, $J^u_LJ^u_R$ and $J^d_LJ^d_R$. The remaining four terms contains up-down mixing due to $J^u_LJ^d_L$, $J^u_RJ^d_R$, $J^u_LJ^d_R$ and $J^d_LJ^u_R$. Collecting all contributions, the Fierz-rearranged geometrical interaction is given by
\begin{align}\label{Fierz}
\mathscr L_{4\psi}^q
=
-\frac12
\Bigg\{
&
(\Lambda_L^Q)_{AB}
(\Lambda_L^Q)_{CD}
(\bar u_{AL}\gamma^a u_{DL})
(\bar u_{CL}\gamma_a u_{BL})
+
(V^\dagger\Lambda_L^QV)_{AB}
(V^\dagger\Lambda_L^QV)_{CD}
(\bar d_{AL}\gamma^a d_{DL})
(\bar d_{CL}\gamma_a d_{BL})
\nonumber\\
&
+
(\Lambda_R^u)_{AB}
(\Lambda_R^u)_{CD}
(\bar u_{AR}\gamma^a u_{DR})
(\bar u_{CR}\gamma_a u_{BR})
+
(\Lambda_R^d)_{AB}
(\Lambda_R^d)_{CD}
(\bar d_{AR}\gamma^a d_{DR})
(\bar d_{CR}\gamma_a d_{BR})
\nonumber\\
&
+
2(\Lambda_L^Q)_{AB}
(V^\dagger\Lambda_L^QV)_{CD}
(\bar u_{AL}\gamma^a d_{DL})
(\bar d_{CL}\gamma_a u_{BL})
+
2(\Lambda_R^u)_{AB}
(\Lambda_R^d)_{CD}
(\bar u_{AR}\gamma^a d_{DR})
(\bar d_{CR}\gamma_a u_{BR})
\nonumber\\
&
-
4(\Lambda_L^Q)_{AB}
(\Lambda_R^u)_{CD}
(\bar u_{AL}u_{DR})
(\bar u_{CR}u_{BL})
-
4(V^\dagger\Lambda_L^QV)_{AB}
(\Lambda_R^d)_{CD}
(\bar d_{AL}d_{DR})
(\bar d_{CR}d_{BL})
\nonumber\\
&
-
4(\Lambda_L^Q)_{AB}
(\Lambda_R^d)_{CD}
(\bar u_{AL}d_{DR})
(\bar d_{CR}u_{BL})
-
4(V^\dagger\Lambda_L^QV)_{AB}
(\Lambda_R^u)_{CD}
(\bar d_{AL}u_{DR})
(\bar u_{CR}d_{BL})
\Bigg\}.
\end{align}
We see that there is only one term in Eq.~\ref{Fierz} that possesses the same structure as the SM charged-current interaction and therefore deserves special attention. If we integrate out the $W^\pm$ boson from the SM charged-current interaction, we get the familiar Fermi interaction
\begin{equation}
\mathscr L_1
=
-\frac{G_F}{\sqrt2}V_{AB}V^*_{CD}
\left(
\bar u_A\gamma^a P_Ld_B
\right)
\left(
\bar d_C\gamma_a P_Lu_D
\right).
\label{SM-Fermi}
\end{equation}
We also have the following contribution from the geometrical interaction 
\begin{equation}
\mathscr L_{2}
=-
G_{ABCD}
(\bar u_A\gamma^a P_Ld_B)
(\bar d_C\gamma_a P_Lu_D),
\end{equation}
where we have defined the flavor tensor
\begin{equation}
G_{ABCD}
=
(\Lambda_L^Q)_{AD}
(V^\dagger\Lambda_L^QV)_{CB}.
\label{FlavorTensor}
\end{equation}
The effective charged-current interaction at energies below the electroweak scale is therefore given by the sum of the Standard Model Fermi interaction and the torsion-induced contribution,
\begin{equation}
\mathscr L_{\rm eff}
=
-\left(\frac{G_F}{\sqrt2}V_{AB}V^*_{CD}+G_{ABCD}\right)
(\bar u_A\gamma^a P_Ld_B)
(\bar d_C\gamma_a P_Lu_D).
\label{EffectiveCC}
\end{equation}
We may write the coefficient of the quark currents in Eq.~\ref{EffectiveCC} in terms of a modified mixing matrix $\hat{V}$ as
\begin{equation}
\frac{G_F}{\sqrt2}\hat{V}_{AB}\hat{V}^*_{CD}=\frac{G_F}{\sqrt2}V_{AB}V^*_{CD}+G_{ABCD}
\end{equation}
Therefore, we have the modified Jarlskog invariant 
\begin{equation}
\hat{J}_{\rm CKM}
=
{\rm Im}
\left(
\hat{V}_{11}\hat{V}_{22}\hat{V}^*_{12}\hat{V}^*_{21}
\right)
={\rm Im}\left[
\left(V_{11}V^*_{21}+\frac{\sqrt2}{G_F}G_{1121}\right)\left(V_{22}V^*_{12}+\frac{\sqrt2}{G_F}G_{2212}\right)
\right].
\end{equation}
In the case of alignment where $\Lambda_L^Q=\operatorname{diag}\left(\lambda_{1L}^Q,\lambda_{2L}^Q,\lambda_{3L}^Q\right)$ and defining $W\equiv V^\dagger\Lambda_L^Q V$, we have
\begin{align}
\hat{J}_{\rm CKM}&=\operatorname{Im}\left[\left(V_{11}V_{21}^*+\frac{\sqrt{2}}{G_F}\lambda_{1L}^QW_{21}\right)
\left(V_{22}V_{12}^*+\frac{\sqrt{2}}{G_F}\lambda_{2L}^QW_{21}^*\right)\right]\nonumber\\
&=J_{\rm CKM}+\frac{\sqrt{2}}{G_F}\operatorname{Im}\left[\lambda_{2L}^QV_{11}V_{21}^*W_{21}^*+\lambda_{1L}^QV_{22}V_{12}^*W_{21}\right],
\end{align}
where
\begin{equation}
J_{\rm CKM}=\operatorname{Im}\left(V_{11}V_{21}^*V_{22}V_{12}^*\right)=c_{12}c_{23}c_{13}^2s_{12}s_{23}s_{13}\sin\delta.
\end{equation}
The \(21\) element of \(X\) is
\begin{align}
W_{21}
&=
\left(V^\dagger\Lambda_L^Q V\right)_{21}
=
\sum_{A=1}^{3}
V_{A2}^*
\lambda_{AL}^Q
V_{A1}
=
\lambda_{1L}^Q V_{12}^*V_{11}
+
\lambda_{2L}^Q V_{22}^*V_{21}
+
\lambda_{3L}^Q V_{32}^*V_{31}.
\end{align}
Using the standard parametrization of \(V\), we have
\begin{align}
V_{12}^*V_{11}
&=
c_{12}s_{12}c_{13}^2,
\\
V_{22}^*V_{21}
&=
-c_{12}s_{12}c_{23}^2
+c_{12}s_{12}s_{23}^2s_{13}^2
\nonumber\\
&\quad
-c_{12}^2c_{23}s_{23}s_{13}e^{i\delta}
+s_{12}^2c_{23}s_{23}s_{13}e^{-i\delta},
\\
V_{32}^*V_{31}
&=
-c_{12}s_{12}s_{23}^2
+c_{12}s_{12}c_{23}^2s_{13}^2
\nonumber\\
&\quad
+c_{12}^2c_{23}s_{23}s_{13}e^{i\delta}
-s_{12}^2c_{23}s_{23}s_{13}e^{-i\delta}.
\end{align}
Therefore,
\begin{align}
W_{21}
={}&
c_{12}s_{12}
\Big[
\lambda_{1L}^Q c_{13}^2
-\lambda_{2L}^Q c_{23}^2
+\lambda_{2L}^Q s_{23}^2s_{13}^2
-\lambda_{3L}^Q s_{23}^2
+\lambda_{3L}^Q c_{23}^2s_{13}^2
\Big]
\nonumber\\
&\qquad\qquad\qquad+
c_{23}s_{23}s_{13}
\left(
\lambda_{3L}^Q-\lambda_{2L}^Q
\right)
\left[
c_{12}^2e^{i\delta}
-s_{12}^2e^{-i\delta}
\right].
\end{align}
Since
\begin{equation}
c_{12}^2e^{i\delta}
-s_{12}^2e^{-i\delta}
=
(c_{12}^2-s_{12}^2)\cos\delta
+i\sin\delta,
\end{equation}
the above expression becomes
\begin{equation}
W_{21}=R_{21}+iI_{21},
\end{equation}
where
\begin{align}
R_{21}
={}&
c_{12}s_{12}
\Big[
\lambda_{1L}^Q c_{13}^2
-\lambda_{2L}^Q c_{23}^2
+\lambda_{2L}^Q s_{23}^2s_{13}^2
-\lambda_{3L}^Q s_{23}^2
+\lambda_{3L}^Q c_{23}^2s_{13}^2
\Big]
\nonumber\\
&\qquad\qquad\qquad\qquad+
\left(
\lambda_{3L}^Q-\lambda_{2L}^Q
\right)
c_{23}s_{23}s_{13}
(c_{12}^2-s_{12}^2)\cos\delta
\end{align}
and
\begin{equation}
I_{21}
=
\left(
\lambda_{3L}^Q-\lambda_{2L}^Q
\right)
c_{23}s_{23}s_{13}\sin\delta.
\end{equation}
Since,
\begin{equation}
V_{11}V_{21}^*=-c_{12}s_{12}c_{13}c_{23}-c_{12}^2c_{13}s_{23}s_{13}e^{-i\delta},
\end{equation}
we obtain
\begin{align}
V_{11}V_{21}^*W_{21}^*
={}&
\left[
-c_{12}s_{12}c_{13}c_{23}
-c_{12}^2c_{13}s_{23}s_{13}e^{-i\delta}
\right]
\left(
R_{21}-iI_{21}
\right)
\nonumber\\
={}&
-c_{12}s_{12}c_{13}c_{23}R_{21}
-c_{12}^2c_{13}s_{23}s_{13}R_{21}\cos\delta
\nonumber\\
&+
c_{12}^2c_{13}s_{23}s_{13}I_{21}\sin\delta
\nonumber\\
&+
i\Big[
c_{12}s_{12}c_{13}c_{23}I_{21}
+c_{12}^2c_{13}s_{23}s_{13}R_{21}\sin\delta
\nonumber\\
&\qquad\qquad
+c_{12}^2c_{13}s_{23}s_{13}I_{21}\cos\delta
\Big].
\end{align}

Therefore,
\begin{equation}
\operatorname{Im}
\left(
V_{11}V_{21}^*W_{21}^*
\right)
=
c_{12}s_{12}c_{13}c_{23}I_{21}
+
c_{12}^2c_{13}s_{23}s_{13}
\left(
R_{21}\sin\delta
+
I_{21}\cos\delta
\right).
\end{equation}

Substituting \(R_{21}\) and \(I_{21}\), we find
\begin{align}
\operatorname{Im}
\left(
V_{11}V_{21}^*W_{21}^*
\right)
={}&
c_{12}s_{12}c_{13}c_{23}^2s_{23}s_{13}
\left(
\lambda_{3L}^Q-\lambda_{2L}^Q
\right)
\sin\delta
\nonumber\\
&+
c_{12}^3s_{12}c_{13}s_{23}s_{13}
\Big[
\lambda_{1L}^Q c_{13}^2
-\lambda_{2L}^Q c_{23}^2
+\lambda_{2L}^Q s_{23}^2s_{13}^2
-\lambda_{3L}^Q s_{23}^2
+\lambda_{3L}^Q c_{23}^2s_{13}^2
\Big]\sin\delta
\nonumber\\
&+
c_{12}^2c_{13}c_{23}s_{23}^2s_{13}^2
\left(
\lambda_{3L}^Q-\lambda_{2L}^Q
\right)
\left[
(c_{12}^2-s_{12}^2)\cos\delta
+\cos\delta
\right].
\end{align}

Using
\begin{equation}
(c_{12}^2-s_{12}^2)+1
=
2c_{12}^2,
\end{equation}
the result simplifies to
\begin{align}
\operatorname{Im}
\left(
V_{11}V_{21}^*W_{21}^*
\right)
={}&
c_{12}s_{12}c_{13}c_{23}^2s_{23}s_{13}
\left(
\lambda_{3L}^Q-\lambda_{2L}^Q
\right)
\sin\delta
\nonumber\\
&+
c_{12}^3s_{12}c_{13}s_{23}s_{13}
\Big[
\lambda_{1L}^Q c_{13}^2
-\lambda_{2L}^Q c_{23}^2
+\lambda_{2L}^Q s_{23}^2s_{13}^2
-\lambda_{3L}^Q s_{23}^2
+\lambda_{3L}^Q c_{23}^2s_{13}^2
\Big]\sin\delta
\nonumber\\
&+
2c_{12}^4c_{13}c_{23}s_{23}^2s_{13}^2
\left(
\lambda_{3L}^Q-\lambda_{2L}^Q
\right)
\sin\delta\cos\delta.
\end{align}
We also have
\begin{equation}
V_{22}V_{12}^*W_{21}
=
\left[
c_{12}s_{12}c_{13}c_{23}
-s_{12}^2c_{13}s_{23}s_{13}e^{i\delta}
\right]
\left(R_{21}+iI_{21}\right).
\end{equation}
Thus,
\begin{align}
\operatorname{Im}
\left(
V_{22}V_{12}^*W_{21}
\right)
={}&
c_{12}s_{12}c_{13}c_{23}I_{21}
-
s_{12}^2c_{13}s_{23}s_{13}
\left(
R_{21}\sin\delta
+
I_{21}\cos\delta
\right)
\nonumber\\
={}&
c_{12}s_{12}c_{13}c_{23}^2s_{23}s_{13}
\left(
\lambda_{3L}^Q-\lambda_{2L}^Q
\right)\sin\delta
\nonumber\\
&-
c_{12}s_{12}^3c_{13}s_{23}s_{13}
\Big[
\lambda_{1L}^Q c_{13}^2
-\lambda_{2L}^Q c_{23}^2
+\lambda_{2L}^Q s_{23}^2s_{13}^2
-\lambda_{3L}^Q s_{23}^2
+\lambda_{3L}^Q c_{23}^2s_{13}^2
\Big]\sin\delta
\nonumber\\
&-
2s_{12}^2c_{12}^2c_{13}c_{23}s_{23}^2s_{13}^2
\left(
\lambda_{3L}^Q-\lambda_{2L}^Q
\right)
\sin\delta\cos\delta.
\end{align}
Combining the two terms, we get
\begin{align}
\hat{J}_{\rm CKM}&=J_{\rm CKM}+\frac{\sqrt{2}}{G_F}\left[c_{12}s_{12}c_{13}c_{23}^2s_{23}s_{13}
\left(\lambda_{3L}^Q-\lambda_{2L}^Q\right)\left(\lambda_{1L}^Q+\lambda_{2L}^Q\right)\sin\delta+\left(\lambda_{2L}^Qc^2_{12}-\lambda_{1L}^Qs^2_{12}\right)M\right],    
\end{align}
where
\begin{align}
M&=c_{12}s_{12}c_{13}s_{23}s_{13}
\Big[
\lambda_{1L}^Q c_{13}^2
-\lambda_{2L}^Q c_{23}^2
+\lambda_{2L}^Q s_{23}^2s_{13}^2
-\lambda_{3L}^Q s_{23}^2
+\lambda_{3L}^Q c_{23}^2s_{13}^2
\Big]\sin\delta \nonumber\\
&\qquad\qquad\qquad\qquad\qquad\qquad\qquad\qquad+2c_{12}^2c_{13}c_{23}s_{23}^2s_{13}^2
\left(
\lambda_{3L}^Q-\lambda_{2L}^Q
\right)
\sin\delta\cos\delta.
\end{align}

\section{Summary and Discussion}\label{summary}

In this paper, we have shown that contrary to expectations, spacetime torsion can affect the dynamics of quarks. Specifically, we have shown that the geometrical four-quark interaction induced by spacetime torsion, which is in turn generated by the dynamics of fermions, provides a new mechanism for FCNC processes because of its chiral and non-universal nature. The resulting flavor-changing effects are governed by the structures $\Lambda_L^{u,d}$ and $\Lambda_R^{u,d}$, which are related to the mixing matrices between the flavor and torsion bases. Consequently, both $\Delta F=1$ non-leptonic decays and $\Delta F=2$ meson mixing can receive tree-level contributions. 

As an illustration, the decay $K^+\rightarrow\pi^0\pi^+$ was considered and its effective Hamiltonian derived. Unlike the SM, where the decay is governed predominantly by left-handed charged-current interactions with smaller penguin contributions, the geometrical interaction produces neutral-current four-quark operators with both left- and right-handed chiral structures. An additional feature is that, even in the absence of a third generation, CP violation can arise from the phases in the mixing matrices, in contrast to the SM, where the corresponding CKM matrix would be real.


The impact of the geometrical interaction may be particularly significant in non-leptonic processes that are highly suppressed in the SM. For example, $K_L\rightarrow\pi^0\pi^0, K_L\rightarrow\pi^+\pi^-$, receive only small SM contributions associated with indirect CP violation and loop effects. Tree-level geometrical contributions could therefore substantially modify their decay rates or CP asymmetries. Similarly, rare charm decays such as $D^0\rightarrow\pi^0\pi^0, D^0\rightarrow K^0\bar K^0, D^0\rightarrow\phi\pi^0$, provide sensitive probes because up-sector FCNCs are strongly suppressed by the GIM mechanism. Any enhancement above SM would consequently signal new flavor-changing interactions~\cite{Petrov:2024ujw}.

The interaction also contributes to non-leptonic $B$-meson decays such as $B\rightarrow K\pi, B\rightarrow\phi K, B\rightarrow\pi\pi, B_s\rightarrow\phi\phi$. Although these modes occur in the SM, several are dominated by loop-induced penguin amplitudes. Tree-level geometrical operators can therefore modify branching fractions, direct CP asymmetries, polarization observables, and isospin amplitudes. Precision measurements at experiments such as \textit{LHCb}, \textit{Belle II}, and future flavor facilities can thus provide important constraints. In particular, channels with extremely suppressed SM amplitudes are promising because even a modest geometrical contribution could generate a comparatively large enhancement. The geometrical enhancement of these decays will be discussed elsewhere.

At first glance it may appear strange that the physics of spacetime has some effect on the dynamics of quarks, because usually we associate the Planck scale with the quantum effects of spacetime. However, the Planck scale is the scale at which quantum effects become dominant in gravitational interaction, while we have considered torsion. The Planck scale appears because the classical weak field limit of any theory of gravitation must be the Newtonian theory, in which Newton's constant $G$ is the coupling constant. On the other hand, the geometrical four-fermion interaction, because it arises from a coupling of torsion with the spin of fermion fields, does not have a classical limit. So the strength of that interaction cannot be fixed by appealing to classical physics. $G$ appears as an overall multiplicative factor for the interaction term, but the spin-torsion coupling constants determine the strength of the term. For the same reason, it is not correct to say that the interaction term is `suppressed' by the Planck scale. Another way to see that is to consider the interaction term in the context of perturbative quantum field theory. This term is clearly not renormalizable in flat space. However, the way this term arises requires that the spacetime is not actually flat but follows Einstein-Cartan theory. So a one-loop amplitude will necessarily create a curvature at the cutoff scale -- it is not possible to remove the cutoff without a prescription of how to handle Planck scale curvature. In other words, we cannot perturbatively quantize this geometrical four-fermion interaction without knowing how to quantize gravity, only tree level amplitudes can be calculated. That is not sufficient to bound the coupling constants. The best that can be said is that this term should be smaller than the effective four-fermion term at low energies arising from the weak neutral current, otherwise its effect would have been seen by now. For most particles, this sets the bound $\lambda < \sqrt{G_F}$, usually somewhat lower~\cite{Chakraborty:2024zek, Barick:2025ahc, Barick:2025iwn}.  

Overall, flavor physics provides a sensitive probe of the geometrical interaction. A comprehensive study of rare kaon, charm, and $B$-meson decays, together with neutral-meson mixing and CP-violating observables, can constrain the flavor-dependent couplings and mixing matrices more strongly than the single illustrative decay considered here. Finally, we note that the geometrical couplings between neutrinos and matter can affect neutrino oscillations~\cite{Barick:2023yrs, Ghose:2025tgc, Barick:2025ynn, Barick:2025iwn}. In addition, neutrino-nucleon scattering~\cite{Formaggio:2012cpf,Gallagher:2011zza,Mosel:2016cwa} can also be affected by the geometrical interaction due to the torsional couplings between the leptons and quarks. The geometrical contribution to neutrino-nucleon interaction processes may be worth further investigation in light of the quark mixing considered in this paper.

\appendix

\section{CP Violation}\label{CPV}

To see explicitly that the torsional interaction in the quark sector breaks CP symmetry, we need the tranformation of fermion fields under CP given by $\mathcal{(CP)}\psi(x)\mathcal{(CP)}^{-1}=C\psi^*(\tilde{x})$, where $\tilde{x}=(t,-\vec{x})$ and $C$ is the representation-dependent unitary charge conjugation matrix satisfying $C^{-1}\gamma^{\mu}C=-(\gamma^{\mu})^T$. Consider the square of the first term in Eq.~(\ref{geom-action.2}), i.e., the term $(\bar{u}_{L}\gamma^a \Lambda^Q_{L}u_{L})(\bar{u}_{L}\gamma_a \Lambda^Q_{L}u_{L})$. Therefore \begin{align}\label{CP}\mathcal{(CP)}\left(\bar{u}_{L}\gamma^a \Lambda^Q_{L}u_{L}\bar{u}_{L}\gamma_a \Lambda^Q_{L}u_{L}\right)\mathcal{(CP)}^{-1}&=\left((u^{*}_{L})^\dag (\gamma^0\gamma^a)^T\Lambda^Q_{L}u^{*}_{L}\right)\left((u^{*}_{L})^\dag (\gamma^0\gamma_a)^T\Lambda^Q_{L}u^{*}_{L}\right)\notag
\\&=\left(\bar{u}_{L}\gamma^a (\Lambda^Q_{L})^Tu_{L}\right)\left(\bar{u}_{L}\gamma_a (\Lambda^Q_{L})^Tu_{L}\right).\end{align}

For this term to break CP invariance, we require that $(\Lambda^Q_{L})^T=\pm\Lambda^Q_{L}$. Since, $\Lambda^Q_{L}$ is hermitian, this implies that the condition for CP conservation is $\Lambda^Q_{L}=\pm(\Lambda^Q_{L})^*$.

\section{Parameter Counting}\label{counting}
We have 4 unitary mixing matrices $V$, $V_1$, $V_2$ and $V_3$ taking into account the torsional four-fermion interaction and charged current interaction of quarks in the mass basis. In general, a 3$\times$3 unitary matrix has 9 real free parameters: 3 angles and 6 complex phases. Thus the four matrices $V$, $V_1$, $V_2$ and $V_3$ by themselves have 36 free parameters. Out of these, 5 complex phases in $V$ can be eliminated due to a residual $[\text{U}(1)]^6$ symmetry in the quark mass term. Hence, we have in total 31 free parameters consisting of 12 angles and 19 complex phases. In the standard convention, the CKM matrix $V$ containing 3 angles and a single complex is phase given by \begin{equation}V=
\begin{pmatrix}
c_{12} c_{13} & s_{12} c_{13} & s_{13} e^{-i\delta} \\
- s_{12} c_{23} - c_{12} s_{23} s_{13} e^{i\delta} &
c_{12} c_{23} - s_{12} s_{23} s_{13} e^{i\delta} &
s_{23} c_{13} \\
s_{12} s_{23} - c_{12} c_{23} s_{13} e^{i\delta} &
- c_{12} s_{23} - s_{12} c_{23} s_{13} e^{i\delta} &
c_{23} c_{13}
\end{pmatrix},
\end{equation} with the shorthand $c_{ij} \equiv \cos\theta_{ij}$, $s_{ij} \equiv \sin\theta_{ij}$. Similarly, we can parametrize the matrix $V_1$ as \begin{equation}\label{form}V_1=C^{(1)}\tilde{V}_1 D^{(1)},\end{equation} where $\tilde{V}_1$ has the form of the CKM matrix given by \begin{equation}\tilde{V}_1=
\begin{pmatrix}
\cos a_1 \cos b_1 & \sin a_1 \cos b_1 & \sin b_1 e^{-i\delta_1} \\
- \sin a_1 \cos c_1 - \cos a_1 \sin c_1 \sin b_1 e^{i\delta_1} &
\cos a_1 \cos c_1 - \sin a_1 \sin c_1 \sin b_1 e^{i\delta_1} &
\sin c_1 \cos b_1 \\
\sin a_1 \sin c_1 - \cos a_1 \cos c_1 \sin b_1 e^{i\delta_1} &
- \cos a_1 \sin c_1 - \sin a_1 \cos c_1 \sin b_1 e^{i\delta_1} &
\cos c_1 \cos b_1
\end{pmatrix}.
\end{equation} The diagonal matrices $C^{(1)}$, $D^{(1)}$ contain the remaining 5 CP-violating phases, \begin{equation}D^{(1)}=\text{diag}(1,e^{i\alpha^{(1)}_1},e^{i\alpha^{(1)}_2}), \hspace{1cm} C^{(1)}=\text{diag}(e^{i\alpha^{(1)}_3},e^{i\alpha^{(1)}_4},e^{i\alpha^{(1)}_5}).\end{equation} We note however that there are two factors of $V_1$ in the geometrical interaction Eq.~(\ref{geom-action.2}). Since $C^{(1)}$ commutes with $\text{diag}({\lambda}^Q_{1L},{\lambda}^Q_{2L},{\lambda}^Q_{3L})$, we can write 
\begin{align}V_1^{\dag}\begin{pmatrix}
\lambda^Q_{1L} & 0 & 0 \\
0 & \lambda^Q_{2L} & 0 \\
0 & 0 & \lambda^Q_{3L}
\end{pmatrix}V_1 =\begin{pmatrix}
1 & 0 & 0 \\
0 & e^{-i\alpha^{(1)}_1} & 0 \\
0 & 0 & e^{-i\alpha^{(1)}_2}
\end{pmatrix}\tilde{V}_1^{\dag}\begin{pmatrix}
\lambda^Q_{1L} & 0 & 0 \\
0 & \lambda^Q_{2L} & 0 \\
0 & 0 & \lambda^Q_{3L}
\end{pmatrix}\tilde{V}_1\begin{pmatrix}
1 & 0 & 0 \\
0 & e^{i\alpha^{(1)}_1} & 0 \\
0 & 0 & e^{i\alpha^{(1)}_2}
\end{pmatrix}\end{align} 
Therefore, the 3 phases of the matrix $C^{(1)}$ have been eliminated. We can count the parameters in $V_2$ and $V_3$ in the same manner. So, we are left with 22 free parameters for the mixing matrices -- 4 for the CKM matrix $V$ and 6 each, consisting of 3 angles and 3 phases, for $V_1$, $V_2$ and $V_3$. The couplings $\lambda^Q_{1L},\lambda^Q_{2L},\lambda^Q_{3L},\lambda^u_{1R},\lambda^u_{2R},\lambda^u_{3R},\lambda^d_{1R},\lambda^d_{2R},\lambda^d_{3R}$ also constitute 9 free parameters, making a total of 31 free parameters in the model. Hence, we have the matrices $V_1$, $V_2$ and $V_3$ given by
\begin{equation}V_1=\tilde{V}_1 {D}^{(1)}, \qquad V_2=\tilde{V}_2 {D}^{(2)}, \qquad V_3=\tilde{V}_3 {D}^{(3)}\end{equation} 
where \begin{equation}\tilde{V}_k=
\begin{pmatrix}
\cos a_k \cos b_k & \sin a_k \cos b_k & \sin b_k e^{-i\delta_k} \\
- \sin a_k \cos c_k - \cos a_k \sin c_k \sin b_k e^{i\delta_k} &
\cos a_k \cos c_k - \sin a_k \sin c_k \sin b_k e^{i\delta_k} &
\sin c_k \cos b_k \\
\sin a_k \sin c_k - \cos a_k \cos c_k \sin b_k e^{i\delta_k} &
- \cos a_k \sin c_k - \sin a_k \cos c_k \sin b_k e^{i\delta_k} &
\cos c_k \cos b_k
\end{pmatrix},
\end{equation} with $k=1,2,3$ and the diagonal phase matrices \begin{equation}{D}^{(k)}=\text{diag}(1,e^{i\alpha^{(k)}_1},e^{i\alpha^{(k)}_2}).\end{equation} 
If we consider $n$ generations, then the mixing matrices will be $n\times n$ unitary matrices consisting of $n^2$ free parameters by themselves with $\frac{n(n-1)}{2}$ angles and $\frac{n(n+1)}{2}$ phases. Out of these, $(2n-1)$ phases can be discarded in the CKM matrix leaving it with $\frac{(n-1)(n-2)}{2}$ phases. For the matrices $V_1$, $V_2$, and $V_3$, we can follow the procedure above and parametrize them in terms of a CKM-like matrix sandwiched between two diagonal phase matrices, as shown in Eq.(~\ref{form}). The CKM-like matrix has $\frac{n(n-1)}{2}$ angles and $\frac{(n-1)(n-2)}{2}$ phases. The two diagonal phase matrices contain $n$ and $(n-1)$ phases, out of which $n$ phases can be eliminated, leaving a total of $\frac{n(n-1)}{2}$ phases in the matrices $V_1$, $V_2$, and $V_3$.

\section{Operators and Coefficients for $B \to \pi\pi$, $D \to \pi\pi$, $D \to K\bar K$}\label{operators}

In this appendix, we give the operators and Wilson coefficients for some of the $\Delta F=1$ processes other than $K \to \pi\pi$ that have not been discussed earlier.

\subsubsection{$B \to K\pi$} 

%
The effective Hamiltonian for the decay $B \to K\pi$ from the geometrical interaction is 
\begin{equation}
\mathcal H_{\text{eff}}^{B \to K\pi}
=
\sum_{q=u,d}
\Big[
C_{q,LL}^{(2)} \mathcal O_{q,LL}^{(2)}
+
C_{q,LR}^{(2)} \mathcal O_{q,LR}^{(2)}
+
C_{q,RL}^{(2)} \mathcal O_{q,RL}^{(2)}
+
C_{q,RR}^{(2)} \mathcal O_{q,RR}^{(2)}
\Big],
\end{equation}
where 
\begin{align}
\mathcal O_{q,LL}^{(2)} &= (\bar s_L \gamma^a b_L)(\bar q_L \gamma_a q_L), \qquad \mathcal O_{q,LR}^{(2)} = (\bar s_L \gamma^a b_L)(\bar q_R \gamma_a q_R),\\
\mathcal O_{q,RL}^{(2)} &= (\bar s_R \gamma^a b_R)(\bar q_L \gamma_a q_L), \qquad \mathcal O_{q,RR}^{(2)} = (\bar s_R \gamma^a b_R)(\bar q_R \gamma_a q_R),
\end{align} 
and the geometrical Wilson coefficients are given by 
\begin{align}C_{u,LL}^{(2)} &= (\Lambda_L^d)_{23}\,(\Lambda_L^u)_{11}, \qquad C_{d,LL}^{(2)} = \frac{1}{2}(\Lambda_L^d)_{23}\,(\Lambda_L^d)_{11},\\C_{q,LR}^{(2)} &= (\Lambda_L^d)_{23}\,(\Lambda_R^q)_{11}, \qquad C_{q,RL}^{(2)} = (\Lambda_R^d)_{23}\,(\Lambda_L^q)_{11},\\ C_{u,RR}^{(2)} &= (\Lambda_R^d)_{23}\,(\Lambda_R^u)_{11}, \qquad C_{d,RR}^{(2)} = \frac{1}{2}(\Lambda_R^d)_{23}\,(\Lambda_R^d)_{11}.\end{align}

\subsubsection{$B \to \phi K$} 

For the decay $B \to \phi K$, the Hamiltonian from the geometrical interaction is 
\begin{equation}
\mathcal H_{\text{eff}}^{B \to \phi K}
=
C_{LL}^{(3)} \mathcal O_{LL}^{(3)}
+
C_{LR}^{(3)} \mathcal O_{LR}^{(3)}
+
C_{RL}^{(3)} \mathcal O_{RL}^{(3)}
+
C_{RR}^{(3)} \mathcal O_{RR}^{(3)},
\end{equation}
where
\begin{align}
\mathcal O_{LL}^{(3)} &= (\bar s_L \gamma^a b_L)(\bar s_L \gamma_a s_L), \qquad \mathcal O_{LR}^{(3)} = (\bar s_L \gamma^a b_L)(\bar s_R \gamma_a s_R), \\
\mathcal O_{RL}^{(3)} &= (\bar s_R \gamma^a b_R)(\bar s_L \gamma_a s_L), \qquad \mathcal O_{RR}^{(3)} = (\bar s_R \gamma^a b_R)(\bar s_R \gamma_a s_R).
\end{align} 
The coefficients are given by 
\begin{align}C_{LL}^{(3)} &= \frac{1}{2}(\Lambda_L^d)_{23}\,(\Lambda_L^d)_{22}, \qquad C_{LR}^{(3)} = (\Lambda_L^d)_{23}\,(\Lambda_R^d)_{22},\\C_{RL}^{(3)} &= (\Lambda_R^d)_{23}\,(\Lambda_L^d)_{22}, \qquad C_{RR}^{(3)} = \frac{1}{2}(\Lambda_R^d)_{23}(\Lambda_R^d)_{22}.\end{align}
As is usually done, we have suppressed the spectator quarks in the decay processes. For other decays, we can get the corresponding operators by making the following changes in $B \to K\pi$: $s \to d$ for $B \to \pi\pi$ and $s \to u$, $b \to c$ for $D \to \pi\pi$. The operators for $D \to K\bar K$ are the ones we get by changing $s \to u$, $b \to c$ in the expressions corresponding to $B \to \phi K$. 
The operators and Wilson coefficients corresponding to $B \to \pi\pi$, $D \to \pi\pi$ and $D \to K\bar K$ are given below.

\subsubsection{$B \to \pi\pi$}  
In this case, we have
\begin{equation}
\mathcal H_{\text{eff}}^{B \to \pi\pi}
=
\sum_{q=u,d}
\Big[
C_{q,LL}^{(4)} \mathcal O_{q,LL}^{(4)}
+
C_{q,LR}^{(4)} \mathcal O_{q,LR}^{(4)}
+
C_{q,RL}^{(4)} \mathcal O_{q,RL}^{(4)}
+
C_{q,RR}^{(4)} \mathcal O_{q,RR}^{(4)}
\Big],
\end{equation}
where
\begin{align}
\mathcal O_{q,LL}^{(4)} &= (\bar d_L \gamma^a b_L)(\bar q_L \gamma_a q_L), \qquad \mathcal O_{q,LR}^{(4)} = (\bar d_L \gamma^a b_L)(\bar q_R \gamma_a q_R), \\
\mathcal O_{q,RL}^{(4)} &= (\bar d_R \gamma^a b_R)(\bar q_L \gamma_a q_L), \qquad \mathcal O_{q,RR}^{(4)} = (\bar d_R \gamma^a b_R)(\bar q_R \gamma_a q_R),\end{align} and
\begin{align}
C_{u,LL}^{(4)} &= (\Lambda_L^d)_{13}\,(\Lambda_L^u)_{11}, \qquad C_{d,LL}^{(4)} = \frac{1}{2}(\Lambda_L^d)_{13}\,(\Lambda_L^d)_{11},\\    
C_{q,LR}^{(4)} &= (\Lambda_L^d)_{13}\,(\Lambda_R^q)_{11}, \qquad C_{q,RL}^{(4)} = (\Lambda_R^d)_{13}\,(\Lambda_L^q)_{11},\\    
C_{u,RR}^{(4)} &= (\Lambda_R^d)_{13}\,(\Lambda_R^u)_{11}, \qquad C_{d,RR}^{(4)} = \frac{1}{2}(\Lambda_R^d)_{13}\,(\Lambda_R^d)_{11}.    
\end{align}

\subsubsection{$D \to \pi\pi$}  
The corresponding effective Hamiltonian is
\begin{equation}
\mathcal H_{\text{eff}}^{D \to \pi\pi}
=
\sum_{q=u,d}
\Big[
C_{q,LL}^{(5)} \mathcal O_{q,LL}^{(5)}
+
C_{q,LR}^{(5)} \mathcal O_{q,LR}^{(5)}
+
C_{q,RL}^{(5)} \mathcal O_{q,RL}^{(5)}
+
C_{q,RR}^{(5)} \mathcal O_{q,RR}^{(5)}
\Big],
\end{equation}
where
\begin{align}
\mathcal O_{q,LL}^{(5)} &= (\bar u_L \gamma^a c_L)(\bar q_L \gamma_a q_L), \qquad \mathcal O_{q,LR}^{(5)} = (\bar u_L \gamma^a c_L)(\bar q_R \gamma_a q_R), \\
\mathcal O_{q,RL}^{(5)} &= (\bar u_R \gamma^a c_R)(\bar q_L \gamma_a q_L), \qquad \mathcal O_{q,RR}^{(5)} = (\bar u_R \gamma^a c_R)(\bar q_R \gamma_a q_R),
\end{align} with coefficients
\begin{align}
C_{u,LL}^{(5)} &= \frac{1}{2}(\Lambda_L^u)_{12}\,(\Lambda_L^u)_{11}, \qquad C_{d,LL}^{(5)} = (\Lambda_L^u)_{12}\,(\Lambda_L^d)_{11},\\    
C_{q,LR}^{(5)} &= (\Lambda_L^u)_{12}\,(\Lambda_R^q)_{11}, \qquad C_{q,RL}^{(5)} = (\Lambda_R^u)_{12}\,(\Lambda_L^q)_{11},\\    
C_{u,RR}^{(5)} &= \frac{1}{2}(\Lambda_R^u)_{12}\,(\Lambda_R^u)_{11}, \qquad C_{d,RR}^{(5)} = (\Lambda_R^u)_{12}\,(\Lambda_R^d)_{11}.    
\end{align}

\subsubsection{$D \to K\bar K$}

The effective Hamiltonian is 
\begin{equation}
\mathcal H_{\text{eff}}^{D \to K\bar K}
=
C_{LL}^{(6)} \mathcal O_{LL}^{(6)}
+
C_{LR}^{(6)} \mathcal O_{LR}^{(6)}
+
C_{RL}^{(6)} \mathcal O_{RL}^{(6)}
+
C_{RR}^{(6)} \mathcal O_{RR}^{(6)},
\end{equation}
where 
\begin{align}
\mathcal O_{LL}^{(6)} &= (\bar u_L \gamma^a c_L)(\bar s_L \gamma_a s_L), \qquad \mathcal O_{LR}^{(6)} = (\bar u_L \gamma^a c_L)(\bar s_R \gamma_a s_R), \\
\mathcal O_{RL}^{(6)} &= (\bar u_R \gamma^a c_R)(\bar s_L \gamma_a s_L), \qquad \mathcal O_{RR}^{(6)} = (\bar u_R \gamma^a c_R)(\bar s_R \gamma_a s_R),
\end{align} and
\begin{align}
C_{LL}^{(6)} &= (\Lambda_L^u)_{12}\,(\Lambda_L^d)_{22}, \qquad C_{LR}^{(6)} = (\Lambda_L^u)_{12}\,(\Lambda_R^d)_{22},\\    
C_{RL}^{(6)} &= (\Lambda_R^u)_{12}\,(\Lambda_L^d)_{22}, \qquad C_{RR}^{(6)} = (\Lambda_R^u)_{12}\,(\Lambda_R^d)_{22}.    
\end{align}


\section{Rephasing Invariants}\label{invariants}
In this appendix, we give the explicit expression of one of the rephasing invariants $\operatorname{Im}(S_1^{1121})$ mentioned in Sec.~\ref{Jarlskog}, containing the phase $\alpha_1^{(1)}$. The other mixed invariants have similar forms.
\begin{align}
\operatorname{Im}(S_1^{1121})
={}&
Z_{11}
\Bigg[
\left(
-c_{12}\,c_{13}\,s_{12}\,c_{23}
-c_{13}\,s_{13}\,s_{23}\,c^2_{12}\,\cos\delta
\right)\left(
I_1\cos\alpha_1^{(1)}
-R_1\sin\alpha_1^{(1)}
\right)\nonumber
\\
&\qquad\qquad\qquad\qquad\qquad
+c_{13}\,s_{13}\,s_{23}\,c^2_{12}\,\sin\delta
\left(
R_1\cos\alpha_1^{(1)}
+I_1\sin\alpha_1^{(1)}
\right)
\Bigg],
\end{align}
where
\begin{align}
R_1
={}&
\lambda^Q_{1L}\sin a_1\cos^2b_1
\nonumber\\
&-\lambda^Q_{2L}\left(\sin a_1\cos a_1\cos c_1+\sin c_1\sin^2a_1\sin b_1\cos\delta_1\right.
\nonumber\\
&+\left.\sin c_1\cos^2a_1\sin^2b_1+\sin a_1\cos a_1\cos c_1
\sin b_1\cos\delta_1\right)
\nonumber\\
&+\lambda^Q_{3L}\left(\sin a_1\cos a_1\sin c_1+\cos c_1\sin b_1\cos\delta_1
+\sin a_1\cos a_1\sin c_1\sin^2b_1\right),
\\
I_1
={}&
\sin b_1\sin\delta_1
\left[
\lambda^Q_{2L}\sin c_1\sin^2a_1
-\lambda^Q_{2L}\sin a_1\cos a_1\cos c_1
+\lambda^Q_{3L}\cos c_1\cos(2a_1)
\right],
\\
Z_{11}
={}&
\lambda^d_{1R}\cos^2a_3\cos^2b_3
\nonumber\\
&+\lambda^d_{2R}
\left[
\sin^2a_3\cos^2c_3
+\cos^2a_3\sin^2c_3\sin^2b_3
+2\sin a_3\cos a_3\cos c_3\sin c_3
\sin b_3\cos\delta_3
\right]
\nonumber\\
&+\lambda^d_{3R}
\left[
\sin^2a_3\sin^2c_3
+\cos^2a_3\cos^2c_3\sin^2b_3
-2\sin a_3\cos a_3\sin c_3\cos c_3
\sin b_3\cos\delta_3
\right].
\end{align}

\bibliographystyle{unsrt}
\bibliography{references_1}

\begin{thebibliography}{10}

\bibitem{Riotto:1999yt}
Antonio Riotto and Mark Trodden.
\newblock {Recent progress in baryogenesis}.
\newblock {\em Ann. Rev. Nucl. Part. Sci.}, 49:35--75, 1999.

\bibitem{Trodden:1998ym}
Mark Trodden.
\newblock {Electroweak baryogenesis}.
\newblock {\em Rev. Mod. Phys.}, 71:1463--1500, 1999.

\bibitem{Dine:2003ax}
Michael Dine and Alexander Kusenko.
\newblock {The Origin of the matter - antimatter asymmetry}.
\newblock {\em Rev. Mod. Phys.}, 76:1, 2003.

\bibitem{Morrissey:2012db}
David~E. Morrissey and Michael~J. Ramsey-Musolf.
\newblock {Electroweak baryogenesis}.
\newblock {\em New J. Phys.}, 14:125003, 2012.

\bibitem{Canetti:2012zc}
Laurent Canetti, Marco Drewes, and Mikhail Shaposhnikov.
\newblock {Matter and Antimatter in the Universe}.
\newblock {\em New J. Phys.}, 14:095012, 2012.

\bibitem{Carena:1997gx}
Marcela Carena, M.~Quiros, A.~Riotto, I.~Vilja, and C.~E.~M. Wagner.
\newblock {Electroweak baryogenesis and low-energy supersymmetry}.
\newblock {\em Nucl. Phys. B}, 503:387--404, 1997.

\bibitem{Akhmedov:1998qx}
Evgeny~K. Akhmedov, V.~A. Rubakov, and A.~Yu. Smirnov.
\newblock {Baryogenesis via neutrino oscillations}.
\newblock {\em Phys. Rev. Lett.}, 81:1359--1362, 1998.

\bibitem{Cline:2011mm}
James~M. Cline, Kimmo Kainulainen, and Michael Trott.
\newblock {Electroweak Baryogenesis in Two Higgs Doublet Models and B meson anomalies}.
\newblock {\em JHEP}, 11:089, 2011.

\bibitem{Enomoto:2021dkl}
Kazuki Enomoto, Shinya Kanemura, and Yushi Mura.
\newblock {Electroweak baryogenesis in aligned two Higgs doublet models}.
\newblock {\em JHEP}, 01:104, 2022.

\bibitem{Fukuda:2024pkh}
Hajime Fukuda, Kohei Kamada, and Thanaporn Sichanugrist.
\newblock {Leptogenesis from magnetic helicity of gauged U(1)$_{B-L}$}.
\newblock {\em JHEP}, 02:192, 2025.

\bibitem{Cartan:1923zea}
E.~Cartan.
\newblock {Sur les vari{\'e}t{\'e}s {\`a} connexion affine et la th{\'e}orie de la relativit{\'e} g{\'e}n{\'e}ralis{\'e}e. (premi{\`e}re partie)}.
\newblock {\em Annales Sci. Ecole Norm. Sup.}, 40:325--412, 1923.

\bibitem{Cartan:1924yea}
E.~Cartan.
\newblock {Sur les vari{\'e}t{\'e}s {\`a} connexion affine et la th{\'e}orie de la relativit{\'e} g{\'e}n{\'e}ralis{\'e}e. (premi{\`e}re partie) (Suite).}
\newblock {\em Annales Sci. Ecole Norm. Sup.}, 41:1--25, 1924.

\bibitem{Gasperini:2013cru}
Maurizio Gasperini.
\newblock {\em {Theory of Gravitational Interactions}}.
\newblock Undergraduate Lecture Notes in Physics. Springer, 2013.

\bibitem{Gursey:1957}
F.~G{\"u}rsey.
\newblock {General relativistic interpretation of some spinor wave equations}.
\newblock {\em Nuovo Cim.}, 5:154, 1957.

\bibitem{Finkelstein:1960}
R.~Finkelstein.
\newblock {Spacetime of the Elementary Particles}.
\newblock {\em J. Math. Phys.}, 1:440, 1960.

\bibitem{Hammond:2002rm}
R.~T. Hammond.
\newblock {Torsion gravity}.
\newblock {\em Rept. Prog. Phys.}, 65:599--649, 2002.

\bibitem{Hehl:1971qi}
F.~W. Hehl and B.~K. Datta.
\newblock {Nonlinear spinor equation and asymmetric connection in general relativity}.
\newblock {\em J. Math. Phys.}, 12:1334--1339, 1971.

\bibitem{Hehl:1974cn}
F.~W. Hehl, G.~D. Kerlick, and P.~Von Der~Heyde.
\newblock {General relativity with spin and torsion and its deviations from einstein's theory}.
\newblock {\em Phys. Rev. D}, 10:1066--1069, 1974.

\bibitem{Hehl:1976kj}
F.~W. Hehl, P.~Von Der~Heyde, G.~D. Kerlick, and J.~M. Nester.
\newblock {General Relativity with Spin and Torsion: Foundations and Prospects}.
\newblock {\em Rev. Mod. Phys.}, 48:393--416, 1976.

\bibitem{Hehl:2007bn}
Friedrich~W. Hehl and Yuri~N. Obukhov.
\newblock {Elie Cartan's torsion in geometry and in field theory, an essay}.
\newblock {\em Annales Fond. Broglie}, 32:157--194, 2007.

\bibitem{Kibble:1961ba}
T.~W.~B. Kibble.
\newblock {Lorentz invariance and the gravitational field}.
\newblock {\em J. Math. Phys.}, 2:212--221, 1961.

\bibitem{Mielke:2017nwt}
Eckehard~W. Mielke.
\newblock {\em {Geometrodynamics of Gauge Fields. On the Geometry of Yang-Mills and Gravitational Gauge Theories}}.
\newblock Mathematical Physics Studies. Springer, 2017.

\bibitem{Sciama:1964wt}
Dennis~W. Sciama.
\newblock {The Physical structure of general relativity}.
\newblock {\em Rev. Mod. Phys.}, 36:463--469, 1964.
\newblock [Erratum: Rev.Mod.Phys. 36, 1103--1103 (1964)].

\bibitem{Poplawski:2009fb}
Nikodem~J. Pop{\l}awski.
\newblock {Classical Physics: Spacetime and Fields}.
\newblock 11 2009.

\bibitem{Chakrabarty:2018ybk}
Subhasish Chakrabarty and Amitabha Lahiri.
\newblock {Different types of torsion and their effect on the dynamics of fields}.
\newblock {\em Eur. Phys. J. Plus}, 133(6):242, 2018.

\bibitem{Hehl:2023khc}
Friedrich~W. Hehl.
\newblock {Four Lectures on Poincar{\'e} Gauge Field Theory}.
\newblock In {\em {International School of Cosmology and Gravitation: Spin, Torsion, Rotation and Supergravity}}, 2023.

\bibitem{Ali:2009ee}
S.~A. Ali, C.~Cafaro, S.~Capozziello, and Ch. Corda.
\newblock {On the Poincare Gauge Theory of Gravitation}.
\newblock {\em Int. J. Theor. Phys.}, 48:3426--3448, 2009.

\bibitem{Chakrabarty:2019cau}
Subhasish Chakrabarty and Amitabha Lahiri.
\newblock {Geometrical contribution to neutrino mass matrix}.
\newblock {\em Eur. Phys. J. C}, 79(8):697, 2019.

\bibitem{Cheng:1984vwu}
Ta-Pei Cheng and Ling-Fong Li.
\newblock {\em {Gauge Theory of Elementary Particle Physics}}.
\newblock Oxford University Press, Oxford, UK, 1984.

\bibitem{Buras:1998raa}
Andrzej~J. Buras.
\newblock {Weak Hamiltonian, CP violation and rare decays}.
\newblock In {\em {Les Houches Summer School in Theoretical Physics, Session 68: Probing the Standard Model of Particle Interactions}}, pages 281--539, 6 1998.

\bibitem{Buras:2020xsm}
Andrzej Buras.
\newblock {\em {Gauge Theory of Weak Decays}}.
\newblock Cambridge University Press, 6 2020.

\bibitem{ParticleDataGroup:2024cfk}
S.~Navas et~al.
\newblock {Review of particle physics}.
\newblock {\em Phys. Rev. D}, 110(3):030001, 2024.

\bibitem{Glashow:1970gm}
S.~L. Glashow, J.~Iliopoulos, and L.~Maiani.
\newblock {Weak Interactions with Lepton-Hadron Symmetry}.
\newblock {\em Phys. Rev. D}, 2:1285--1292, 1970.

\bibitem{Buras:1991jm}
Andrzej~J. Buras, Matthias Jamin, Markus~E. Lautenbacher, and Peter~H. Weisz.
\newblock {Effective Hamiltonians for $\Delta S = 1$ and $\Delta B = 1$ Nonleptonic Decays Beyond the Leading Logarithmic Approximation}.
\newblock {\em Nucl. Phys. B}, 370:69--104, 1992.

\bibitem{Buras:1993dy}
Andrzej~J. Buras, Matthias Jamin, and Markus~E. Lautenbacher.
\newblock {The Anatomy of $\epsilon'/\epsilon$ Beyond Leading Logarithms with Improved Hadronic Matrix Elements}.
\newblock {\em Nucl. Phys. B}, 408:209--285, 1993.

\bibitem{Ciuchini:1993vr}
Marco Ciuchini, Enzo Franco, Guido Martinelli, and Laura Reina.
\newblock {The $\Delta S = 1$ Effective Hamiltonian Including Next-to-Leading Order QCD and QED Corrections}.
\newblock {\em Nucl. Phys. B}, 415:403--462, 1994.

\bibitem{Buchalla:1995vs}
Gerhard Buchalla, Andrzej~J. Buras, and Markus~E. Lautenbacher.
\newblock {Weak Decays Beyond Leading Logarithms}.
\newblock {\em Rev. Mod. Phys.}, 68:1125--1144, 1996.

\bibitem{Bigi:2000yz}
Ikaros~I. Bigi and A.~I. Sanda.
\newblock {\em {CP violation}}, volume~9.
\newblock Cambridge University Press, 9 2009.

\bibitem{Donoghue:2022wrw}
John~F. Donoghue, Eugene Golowich, and Barry~R. Holstein.
\newblock {\em {Dynamics of the Standard Model}: {Second edition}}.
\newblock Cambridge University Press, 11 2022.

\bibitem{Carrasco:2013jda}
Nuria Carrasco, Vittorio Lubicz, and Luca Silvestrini.
\newblock {Vacuum Insertion Approximation and the {\ensuremath{\Delta}}I=1/2 rule: A lattice QCD test of the na{\"\i}ve factorization hypothesis for K , D , B and static mesons}.
\newblock {\em Phys. Lett. B}, 736:174--179, 2014.

\bibitem{Becirevic:2004ya}
D.~Becirevic, G.~Isidori, V.~Lubicz, G.~Martinelli, F.~Mescia, S.~Simula, C.~Tarantino, and G.~Villadoro.
\newblock {The K ---{\ensuremath{>}} pi vector form-factor at zero momentum transfer on the lattice}.
\newblock {\em Nucl. Phys. B}, 705:339--362, 2005.

\bibitem{Belyaev:2000xt}
A.~Belyaev, M.~Chizhov, A.~Dorokhov, John~R. Ellis, M.~E. Gomez, and S.~Lola.
\newblock {Charged lepton flavor violation in kaon decays in supersymmetric theories}.
\newblock {\em Eur. Phys. J. C}, 22:715--726, 2002.

\bibitem{Buras:2014maa}
Andrzej~J. Buras, Jean-Marc G{\'e}rard, and William~A. Bardeen.
\newblock {Large $N$ Approach to Kaon Decays and Mixing 28 Years Later: $\Delta I = 1/2$ Rule, $\hat B_K$ and $\Delta M_K$}.
\newblock {\em Eur. Phys. J. C}, 74:2871, 2014.

\bibitem{Buras:2016fys}
Andrzej~J. Buras and Jean-Marc Gerard.
\newblock {Final state interactions in $K\rightarrow \pi \pi $ decays: $\Delta I=1/2$ rule vs. $\varepsilon '/\varepsilon $}.
\newblock {\em Eur. Phys. J. C}, 77(1):10, 2017.

\bibitem{Cirigliano:2003gt}
V.~Cirigliano, G.~Ecker, H.~Neufeld, and A.~Pich.
\newblock {Isospin breaking in K ---{\ensuremath{>}} pi pi decays}.
\newblock {\em Eur. Phys. J. C}, 33:369--396, 2004.

\bibitem{RBC:2020kdj}
R.~Abbott et~al.
\newblock {Direct CP violation and the $\Delta I=1/2$ rule in $K\to\pi\pi$ decay from the standard model}.
\newblock {\em Phys. Rev. D}, 102(5):054509, 2020.

\bibitem{Jarlskog:1985cw}
C.~Jarlskog.
\newblock {A Basis Independent Formulation of the Connection Between Quark Mass Matrices, CP Violation and Experiment}.
\newblock {\em Z. Phys. C}, 29:491--497, 1985.

\bibitem{Jarlskog:1985ht}
C.~Jarlskog.
\newblock {Commutator of the Quark Mass Matrices in the Standard Electroweak Model and a Measure of Maximal CP Nonconservation}.
\newblock {\em Phys. Rev. Lett.}, 55:1039, 1985.

\bibitem{Branco:1987mj}
G.~C. Branco and L.~Lavoura.
\newblock {Rephasing Invariant Parametrization of the Quark Mixing Matrix}.
\newblock {\em Phys. Lett. B}, 208:123--130, 1988.

\bibitem{Branco:1999fs}
Gustavo Branco, Luís Lavoura, and João Silva.
\newblock {\em CP Violation}.
\newblock Oxford University Press, 07 1999.

\bibitem{Wu:1985ea}
Dan-di Wu.
\newblock {The Rephasing Invariants and CP}.
\newblock {\em Phys. Rev. D}, 33:860, 1986.

\bibitem{Petrov:2024ujw}
Alexey~A. Petrov.
\newblock {Charm physics}.
\newblock {\em Eur. Phys. J. ST}, 233(2):439--456, 2024.

\bibitem{Chakraborty:2024zek}
Arnab Chakraborty and Amitabha Lahiri.
\newblock {Parity nonconservation induced by spacetime geometry}.
\newblock {\em Eur. Phys. J. Plus}, 139(10):955, 2024.

\bibitem{Barick:2025ahc}
Riya Barick and Amitabha Lahiri.
\newblock {Exploring the~Effect of~Chiral Torsion on~Neutrino Oscillation in~Long Baseline Experiments}.
\newblock {\em Springer Proc. Phys.}, 432:1170--1173, 2026.

\bibitem{Barick:2025iwn}
Riya Barick, Indrajit Ghose, Srubabati Goswami, Amitabha Lahiri, and Sushant~K. Raut.
\newblock {Probing geometrical NSI at the DUNE experiment}.
\newblock {\em JHEP}, 07:099, 2026.

\bibitem{Barick:2023yrs}
Riya Barick, Indrajit Ghose, and Amitabha Lahiri.
\newblock {Effect of spacetime geometry on neutrino oscillations}.
\newblock {\em Eur. Phys. J. Plus}, 139(6):461, 2024.

\bibitem{Ghose:2025tgc}
Indrajit Ghose and Amitabha Lahiri.
\newblock {Effects of spacetime geometry on neutrino oscillation inside a core-collapse supernova}.
\newblock {\em JCAP}, 05:020, 2025.

\bibitem{Barick:2025ynn}
Riya Barick and Amitabha Lahiri.
\newblock {Torsional modulation of atmospheric neutrino oscillation}.
\newblock {\em JHEP}, 10:017, 2025.

\bibitem{Formaggio:2012cpf}
J.~A. Formaggio and G.~P. Zeller.
\newblock {From eV to EeV: Neutrino Cross Sections Across Energy Scales}.
\newblock {\em Rev. Mod. Phys.}, 84:1307--1341, 2012.

\bibitem{Gallagher:2011zza}
H.~Gallagher, G.~Garvey, and G.~P. Zeller.
\newblock {Neutrino-nucleus interactions}.
\newblock {\em Ann. Rev. Nucl. Part. Sci.}, 61:355--378, 2011.

\bibitem{Mosel:2016cwa}
Ulrich Mosel.
\newblock {Neutrino Interactions with Nucleons and Nuclei: Importance for Long-Baseline Experiments}.
\newblock {\em Ann. Rev. Nucl. Part. Sci.}, 66:171--195, 2016.

\end{thebibliography}

\end{document}